\documentclass[reprint,NumberedRefs]{JASAmod}
\newlength{\figwidth}
\usepackage{pifont}
\renewcommand{\pi}[0]{\textrm{\Pisymbol{psy}{"70}}} 
\newcommand{\rmu}[0]{\textrm{\Pisymbol{psy}{"6D}}} 

\begin{document}
\title[]{A method for analyzing sampling jitter in audio equipment}
\author{Makoto Takeuchi}
\email{takeuchi@phys.c.u-tokyo.ac.jp}
\author{Haruo Saito}
\affiliation{Graduate School of Arts and Sciences, The University of Tokyo, Tokyo, 153-8902, JAPAN}
\begin{abstract}
A method for analyzing sampling jitter in audio equipment is proposed. The method is based on the time-domain analysis where the time fluctuations of zero-crossing points in recorded sinusoidal waves are employed to characterize jitter. This method enables the separate evaluation of jitter in an audio player from those in audio recorders when the same playback signal is simultaneously fed into two audio recorders. Experiments are conducted using commercially available portable devices with a maximum sampling rate of 192~000 samples per second. The results show jitter values of a few tens of picoseconds can be identified in an audio player. Moreover, the proposed method enables the separation of jitter from phase-independent noise utilizing the left and right channels of the audio equipment. As such, this method is applicable for performance evaluation of audio equipment, signal generators, and clock sources.
\end{abstract}

\maketitle

\section{\label{Section1} Introduction}
Sampling jitter in audio equipment is an error in the sampling instants from the ideal timing, i.e., $t[i]=(i-1)f_\mathrm{S}^{-1}$, where $f_\mathrm{S}$ is the sampling frequency of digital-to-analog converter (DAC) and analog-to-digital converter (ADC). Sampling jitter $j_\mathrm{S}(t)$ causes the sampling instants to be changed to $t[i]=(i-1)f_\mathrm{S}^{-1}+j_\mathrm{S}(t[i])$. Hence, sampling jitter affects the performance of audio equipment. The sampling jitter has conventionally been analyzed in the frequency domain\citep{Dunn92, Dunn94Feb, Dunn94May, Dunn00}. In this method, one plays back a sinusoidal wave whose frequency is $f_\mathrm{P}/4$, where $f_\mathrm{P}$ is the sampling frequency of the audio player, records it, and then examines the frequency response using a window function with small side-lobe levels such as a Blackman Harris window. In addition to frequency-domain analysis (FDA), time domain analysis (TDA) of sampling jitter has been conducted\citep{Nishimura10}. The Hilbert transform has been employed to obtain the real jitter waveform $j_\mathrm{S}(t)$. The advantage of TDA is that one can separately extract jitter and amplitude modulation (AM) from a recorded waveform,which is not possible with FDA.

In this study, we propose an efficient and powerful method to characterize sampling jitter in audio equipment. The proposed method comprises two key elements. The first is an improved TDA termed zero-crossing analysis (ZCA). To apply this method, the zero-crossing points (ZCPs) of a recorded waveform are analyzed, following which the ZCPs of an ideal sinusoidal wave are calculated. Time differences in ZCPs between the recorded waveform and an ideal sinusoidal wave contain the jitter information. We term these time differences ``zero-crossing fluctuations (ZCFs).'' The ZCA enables us to extract jitter information from a recorded waveform even when the input signal contains both jitter and AM. The second key element is the simultaneous recording of the same playback signal with two audio recorders to generate two independent waveforms. We term this ``double recorder setup (DRS).'' Because ZCA preserves absolute time information, we can exactly compare and calculate the positive and negative correlations of ZCFs between the two generated waveforms. Based on the addition rule of probability, the sampling jitter of the player and that of the recorders can be individually evaluated. 

Note that the proposed method requires neither an optional output clock signal synchronized with the recorders' internal clock nor an external clock generator that is more precise than the internal clock. Thus, the proposed method is possible using low-cost recorders and is feasible at an end-user level. The proposed method can be also applied to high-frequency phase noise and jitter measurements. Replacing the two recorders with two digital oscilloscopes with higher sampling rates, we can use the proposed method for performance evaluation of various signal and clock generators that output high-frequency sinusoidal waves. 

The method is somewhat similar to the reciprocal calibration of microphones, which has been used since the 1940s\citep{MacLean40, Barrera-Figueroa18}; however, it does not require the bidirectional use of devices. The DRS is similar to the setup of the cross-spectrum method (CSM) for phase noise measurement\citep{Rubiola10}. In CSM, repeating FDA with double instruments reduces the influence of the instruments to $1/\sqrt{m_\mathrm{CS}}$, where $m_\mathrm{CS}$ is the number of measurements. The proposed method is the so-called TDA version of CSM. The influence of the instruments is canceled by using DRS.

In this study, we focus on the performance of audio equipment; however, it contributes to the field of human audibility\citep{Ashihara05, Melchior19,Nittono20}. Previous jitter studies\citep{Ashihara05} demonstrated that the threshold of perceptual detection of random jitter in music signals is large, but the original jitter, i.e., the one that is already existing before adding extra jitter, has not been controlled at that time. Researchers can quickly select instruments with minimal jitter using the method proposed herein. It provides an opportunity to examine how the detection threshold is reduced after the testees are well-trained using audio players with lower levels of jitter and using music signals with which slight artificial jitter is compounded. Moreover, the proposed method helps to diagnose whether the player is appropriately operating in all audibility studies. 

The remainder of this study is organized as follows. Section \ref{Section2} describes the principles on which the proposed method is based. Section \ref{Section4} presents the experimental procedure. Section \ref{Sec:result_all} presents our results. In Section \ref{Section5}, multiple perspectives are discussed, and Section \ref{Section6} provides a summary and outlook. We performed numerical calculations to confirm the accuracy of the proposed method, the details of which are presented in Appendix.

\section{\label{Section2}Principles}
\subsection{Classification of noise in a playback signal and a recorded waveform}
\begin{figure}[htbp]
\figcolumn{\fig{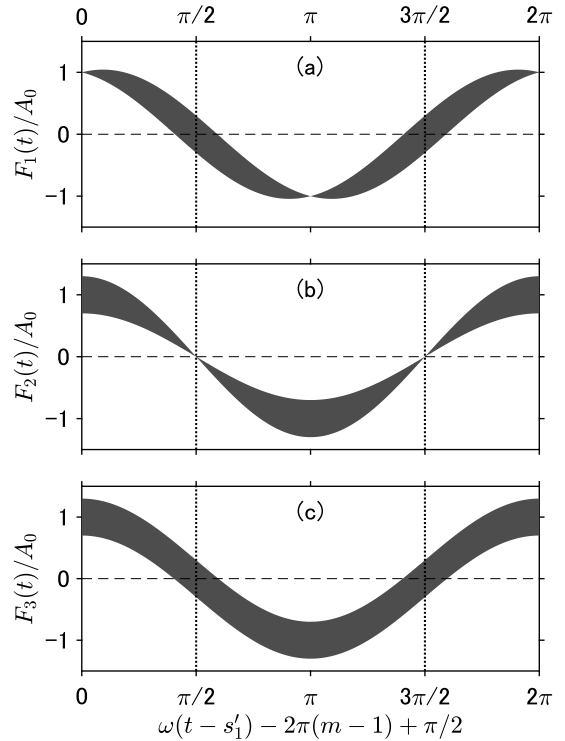}{0.4\figwidth}{}}
\caption{\label{Figure1}Schematic of sinusoidal signals modulated by (a) jitter, (b) AM, and (c) PI noise. The horizontal axis represents the phase of a pure sinusoidal wave. Time $t=s'_1$ is the first zero-crossing time of the pure sinusoidal wave, and $m$ is the number of cycles. The filled bands represent the range of fluctuation during the repetition.}
\end{figure}

A pure sinusoidal wave $F_{\rm pure}(t)$ is expressed as follows: 
\linenomath
\begin{align}
F_\mathrm{pure}(t)=A_0\cos(\omega t+\theta_0), \label{e1}
\end{align}
where $t$, $\omega$, $\theta_0$, and $A_0$ are the time, angular frequency, initial phase, and amplitude of the wave, respectively. When one plays a digital audio file by which a pure sinusoidal wave is expected to be reproduced, the resulting playback signal is not a pure sinusoidal wave. It contains (i) jitter, (ii) AM, and (iii) phase-independent (PI) noise. Here, we explain these three noise patterns individually. (i)~Jitter is the deviation of the playback timing at each point. When jitter is present, $t$ is replaced by $t+j(t)$, where $j(t)$ is the jitter of the player. The pure sinusoidal wave $F_\mathrm{pure}(t)$ is thus changed to 
\linenomath
\begin{align}
F_{1}(t)
&=A_0\cos\{\omega [t+j(t)]+\theta_0\}\\
&= F_\mathrm{pure}(t)+n_\mathrm{jitter}(t). 
\end{align}
When $\omega j(t) \ll 1$,  the noise caused by jitter $n_\mathrm{jitter}(t)$ is expressed as follows: 
\linenomath
\begin{align}
n_\mathrm{jitter}(t)
&\approx-\omega j(t) A_0\sin(\omega t+\theta_0).
\end{align}
This denotes that jitter is prominent when the signal is near zero, i.e., $F_\mathrm{1}(t)\approx 0$, and that it is easier to measure the jitter $j(t)$ when $A_0$ and $\omega$ are larger. A conceptual view of jitter is shown in \autoref{Figure1}(a). 
(ii)~AM is the amplitude variation of a wave concerning time. If AM is present, $A_0$ is replaced by $A_0+A_\mathrm{M}(t)$, where $A_\mathrm{M}(t)$ is a continuous function that represents AM at time $t$. The pure sinusoidal wave $F_\mathrm{pure}(t)$ is changed to 
\linenomath
\begin{align}
F_2(t)
&=[A_0+A_\mathrm{M}(t)]\cos(\omega t+\theta_0)\\
&=F_{\rm pure}(t)+n_\mathrm{AM}(t). \label{e4}
\end{align}
The noise caused by AM $n_\mathrm{AM}(t)$ is expressed as follows: 
\linenomath
\begin{align}
n_{\rm AM}(t)=A_\mathrm{M}(t)\cos(\omega t+\theta_0). \label{e4b}
\end{align}
Consequently, AM increases when $\cos(\omega t+\theta_0)$ increases, in contrast to the behavior of jitter. A conceptual view of AM is shown in \autoref{Figure1}(b). (iii)~The actual wave contains not only jitter and AM but also PI noise, which represents all other types of noise that are not categorized as jitter or AM. If PI noise is present, the pure sinusoidal wave $F_\mathrm{pure}(t)$ is changed to 
\linenomath
\begin{align}
F_3(t)=F_\mathrm{pure}(t)+n_\mathrm{PI}(t), 
\end{align}
where $n_\mathrm{PI}(t)$ denotes PI noise. A conceptual view of PI noise is shown in \autoref{Figure1}(c). Considering all three noise patterns, the total noise $n_\mathrm{total}(t)$ and actual playback signal $c(t)$ become 
\linenomath
\begin{align}
n_\mathrm{total}(t)&=n_{\rm jitter}(t)+n_{\rm AM}(t)+n_\mathrm{PI}(t), \label{e6}\\ 
c(t)&=F_\mathrm{pure}(t)+n_\mathrm{total}(t). \label{Eq:c}
\end{align}
In the actual digital audio players, $n_{\rm jitter}(t)$ comes primarily from the internal clock module, and $n_\mathrm{AM}(t)$ is from DAC units. The output amplifier of the DAC contributes to $n_\mathrm{PI}(t)$. 

In the proposed method, we focus on the ZCPs of impure sinusoidal waves. The ZCPs of $F_\mathrm{pure}(t)$, represented as $s'_k$, satisfy $F_\mathrm{pure}(s'_k)=0$; thus, they are given by 
\linenomath
\begin{align}
s'_k=\frac{1}{\omega}\left(k\pi-\frac{\pi}{2}-\theta_0\right),  \label{Eq:s'_k}
\end{align}
where $k$ is the index of ZCPs. At $t=s_k'$, the playback signal $c(t)$ is not equal to zero and is given by 
\linenomath
\begin{align}
c(s'_k)= (-1)^k\omega A_0 j(s'_k) +n_\mathrm{PI}(s'_k), \label{Eq:F_actual}
\end{align}
which can be derived from Eq. (\ref{Eq:c}). 

The aforementioned noise classification is also appropriate in the recording process. Although a data array is a set of discrete variables rather than a continuous function, we refer to it as a ``waveform'' in the following. When one records a pure sinusoidal playback signal, the recorded waveform is not equal to the sampled data points of a pure sinusoidal wave because it contains (i) jitter, (ii) AM, and (iii) PI noise. In real digital audio recorders, jitter originates primarily from the internal clock module, whereas AM originates from ADC units. The driver amplifier for ADC contributes to PI noise. 

As demonstrated later in this study, PI noise becomes comparable to jitter in the case of recent audio equipment with small jitter values, typically less than $100~\mathrm{ps}$. Jitter and PI noises can be separated using the left and right channels of the player and recorder. 

\subsection{\label{Sec:player}Modeling of digital audio player}
\begin{figure}[htbp]
\figcolumn{\fig{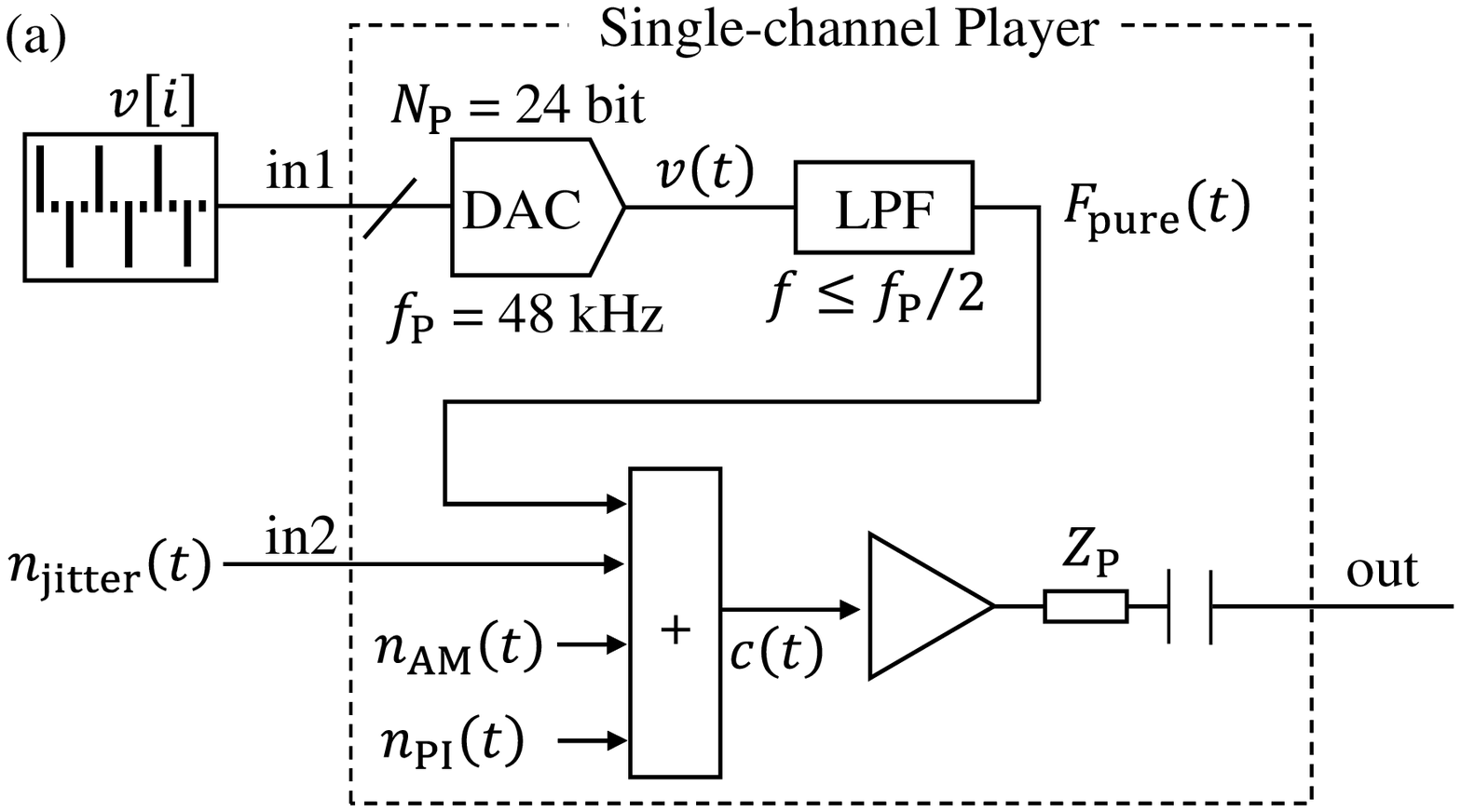}{0.45\figwidth}{}
\fig{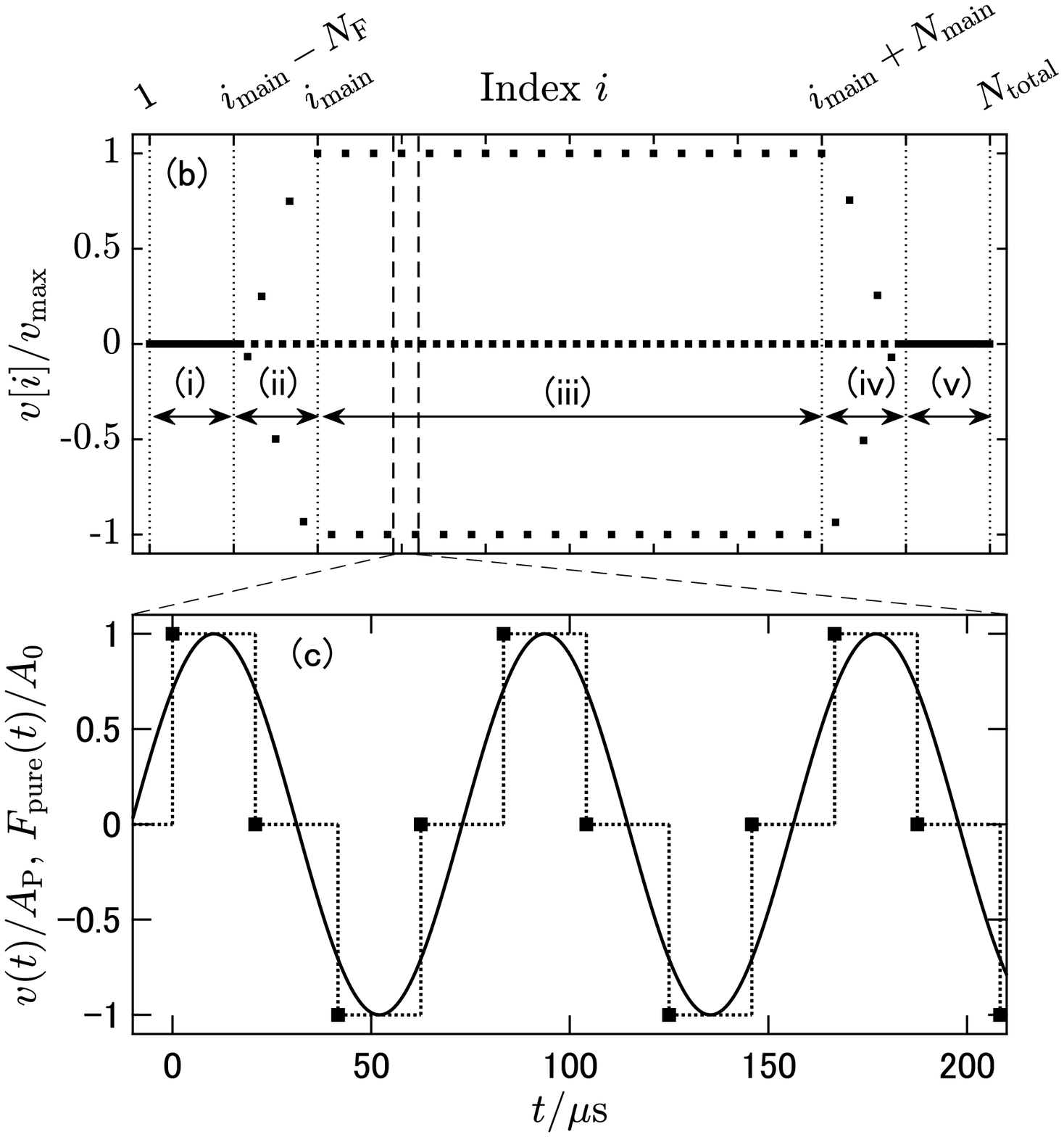}{0.45\figwidth}{}
\fig{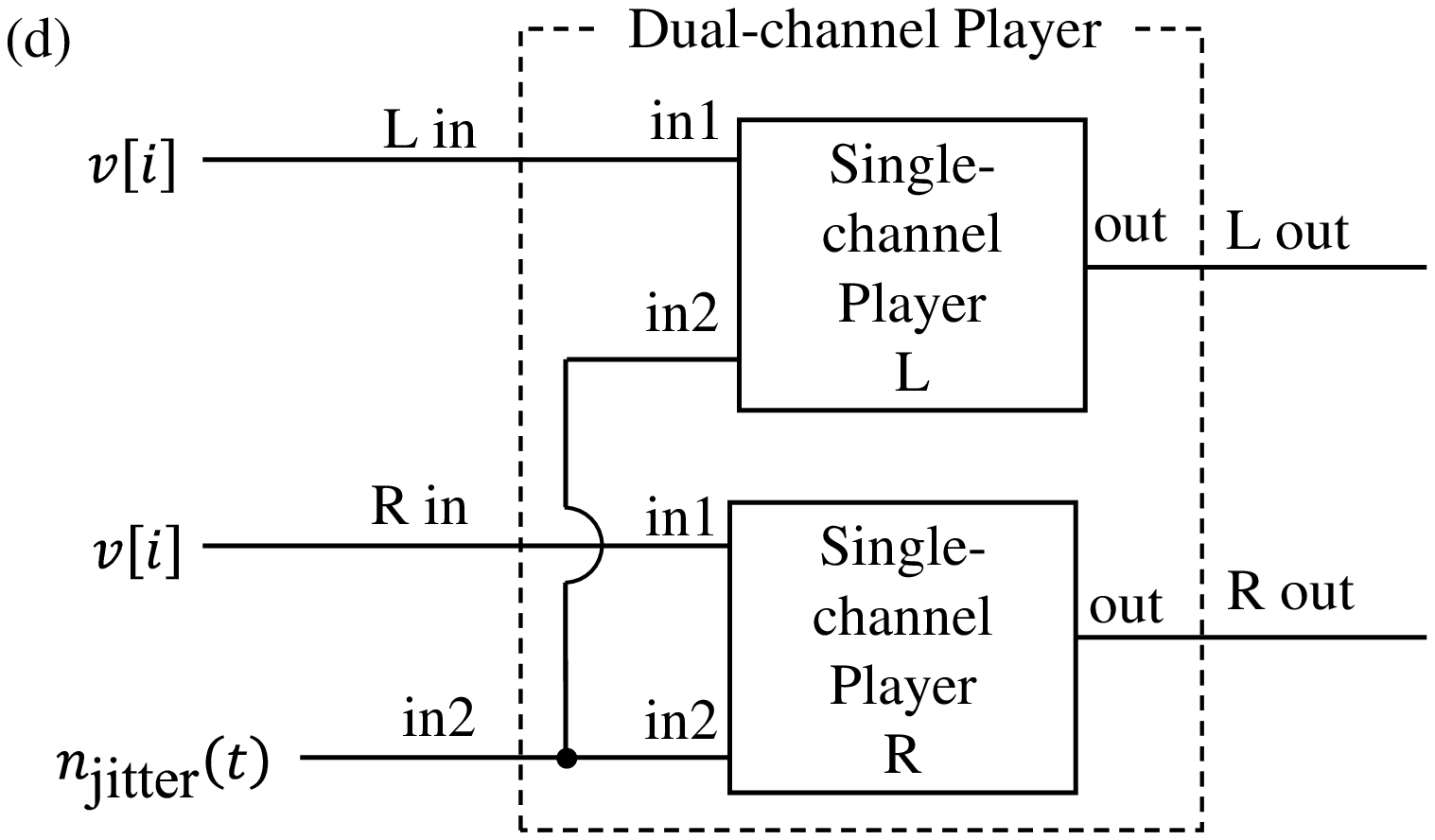}{0.45\figwidth}{}}
\caption{\label{Fig:player}{(a) Model diagram of single-channel digital audio player. (b) Playback waveform; dots have been reduced to improve visibility. (c) Relationships among DAC output $v(t)$ (dotted line), signal after LPF $F_\mathrm{pure}(t)$ (solid line), and playback waveform $v_i$ (black square). (d) Model diagram demonstrating a dual-channel digital audio player.}}
\end{figure}

Herein, we represent how jitter, AM, and PI noises, introduced in the previous section, can be realized in a real playback process. Hence, we introduce a model of a digital audio player. A schematic of a single-channel digital audio player is shown in \autoref{Fig:player}(a). 
The parameters are fixed to be the same as those of the experimental conditions described in Section \ref{Section4} and can be arbitrarily selected depending on the experimental conditions. The DAC in the player, assumed to be ideal and noise-free, performs the conversion at $N_\mathrm{P}$ bits with a sampling rate $f_\mathrm{P}$. In this study, $N_\mathrm{P}$ and $f_\mathrm{P}$ are set to $24~\mathrm{bit}$ and $48~\mathrm{kHz}$, respectively. 

The playback waveform is represented as $v[i]$, where $i$ is a natural number, and $v[i]$ is a $N_\mathrm{P}$ bit signed integer. The length of $v[i]$ is $N_\mathrm{total}$. The waveform $v[i]$ used in this study is shown in \autoref{Fig:player}(b). It can be separated into five parts, labeled (i), (ii), (iii), (iv), and (v) as shown in the figure. Dots have been reduced to improve visibility in the figure. The horizontal axis represents index $i$. (i) Silent part, where the playback waveform is $v[i]=0$ for $1\leqslant i \leqslant i_\mathrm{main}-N_\mathrm{F}-1$. As shown below, $i_\mathrm{main}$ is the first index of the main part, and $N_\mathrm{F}$ is the length of the fade part. We set $i_\mathrm{main}=480~000$ and $N_\mathrm{F}=240~000$. Therefore, the temporal duration of the silent part becomes $(i_\mathrm{main}-N_\mathrm{F})f_\mathrm{P}^{-1}\approx 5~\mathrm{s}$. (ii) Fade-in part: the playback waveform has a form of 
\linenomath
\begin{align}
v[i]=
&\left\{v_\mathrm{min}+\left[1+\cos\left(\pi\frac{i-i_\mathrm{main}}{N_\mathrm{F}}\right)\right]\right.\nonumber\\
&\left. \quad \times \frac{v_\mathrm{max}-v_\mathrm{min}}{2}\right\}\cos\left(2\pi\frac{\mathrm{mod}[i-i_\mathrm{main},4]}{4}\right)\label{Eq:fade-in}
\end{align}
in the region of $i_\mathrm{main}-N_\mathrm{F} \leqslant i \leqslant i_\mathrm{main}-1$, where $v_\mathrm{max}:= 2^{N_\mathrm{P}-1}-1=8~388~607$ is the maximum value of a $N_\mathrm{P}$ bit signed integer. The initial amplitude in the fade-in part $v_\mathrm{min}$ is set to $256$. The temporal duration of the fade-in part becomes $N_\mathrm{F}f_\mathrm{P}^{-1}\approx 5~\mathrm{s}$. (iii) Main part: the playback waveform is a repetition of $(v_\mathrm{max}, 0, -v_\mathrm{max}, 0)$ for $i_\mathrm{main}\leqslant i \leqslant i_\mathrm{main}+N_\mathrm{main}-1$, i.e., the playback waveform in the main part is expressed as follows: 
\linenomath
\begin{align}
v[i]=v_\mathrm{max}\cos\left(2\pi\frac{\mathrm{mod}[i-i_\mathrm{main},4]}{4}\right). \label{Eq:v_i}
\end{align}
In this study, we set $N_\mathrm{main}=1~440~000$. Therefore, the main part of the waveform begins at $(i_\mathrm{main})f_\mathrm{P}^{-1}\approx 10~\mathrm{s}$ after playback is started, and continues $(N_\mathrm{main})f_\mathrm{P}^{-1}\approx 30~\mathrm{s}$. (iv) Fade-out part: this part begins after the main part and its length is the same as that of the fade-in part. The waveform of the fade-out part is the reversed sequence of the fade-in part, which is expressed as Eq. (\ref{Eq:fade-in}). (v) Second silent part: this part begins after the fade-out part and its length is the same as that of the first silent part. Consequently, the total length of the playback signal, which is the summation of the length of its five parts, is $N_\mathrm{total} f_\mathrm{P}^{-1} = (2i_\mathrm{main}+N_\mathrm{main})f_\mathrm{P}^{-1}\approx 50~\mathrm{s}$. 

The DAC output voltage, represented as $v(t)$, is a square wave. For $t_\mathrm{P}+(i-1)f_\mathrm{P}^{-1} \leqslant t < t_\mathrm{P}+i f_\mathrm{P}^{-1}$, $v(t)$ is expressed as $v(t)=A_\mathrm{P} v[i]/v_\mathrm{max}$. The time when playback is started is $t_\mathrm{P}$. In this section, we set $t_\mathrm{P}=-(i_\mathrm{main}+N_\mathrm{main}/6)f_\mathrm{P}^{-1}\approx -15~\mathrm{s}$.  

A low-pass filter (LPF) is connected after the DAC. The cut-off frequency of the LPF is assumed to be $f_\mathrm{P}/2$. Because the square wave $v(t)$ is smoothed by the LPF, the output voltage after the LPF becomes a pure sinusoidal wave expressed as Eq. (\ref{e1}). The frequency of the pure sinusoidal wave becomes $f_\mathrm{C}:=\omega/2\pi=f_\mathrm{P}/4$ since the main part of the playback waveform is expressed as Eq. (\ref{Eq:v_i}). The relationship between $v(t)$ and $F_\mathrm{pure}(t)$ is depicted in \autoref{Fig:player}(c). The dotted and solid lines represent $v(t)$ and $F_\mathrm{pure}(t)$, respectively. For comparison with $v(t)$, the playback waveform $v[i]$ is plotted as  black squares at the position $t=t_\mathrm{P}+(i-1)f_\mathrm{P}^{-1}$ and $v(t)=A_\mathrm{P}v[i]/v_\mathrm{max}$. As mentioned above, the components of $3f_\mathrm{C}, 5f_\mathrm{C},\cdots$ in $v(t)$ are perfectly attenuated by the LPF, and a pure sinusoidal wave of frequency $f_\mathrm{C}$ remains.

The playback signal $c(t)$ is obtained by adding $n_\mathrm{jitter}(t)$, $n_\mathrm{AM}(t)$, $n_\mathrm{PI}(t)$ to $F_\mathrm{pure}(t)$, as shown in Eqs. (\ref{e6}) and (\ref{Eq:c}). In real audio player, jitter is primarily caused by fluctuation of $f_\mathrm{P}$. However, in this model, DAC and LPF are ideal, and jitter noise is added after the LPF. The output is made by a buffer amplifier with a direct current (DC)-blocking capacitor.
 
A model diagram of a dual-channel digital audio player is presented in \autoref{Fig:player}(d). It comprises two single-channel digital audio players, the jitter inputs of which are assumed to be equipotential. This model diagram is used in Section \ref{Sec:jitter-PI}.

\subsection{Modeling of digital audio recorder}
\begin{figure}[htbp]
\figcolumn{\fig{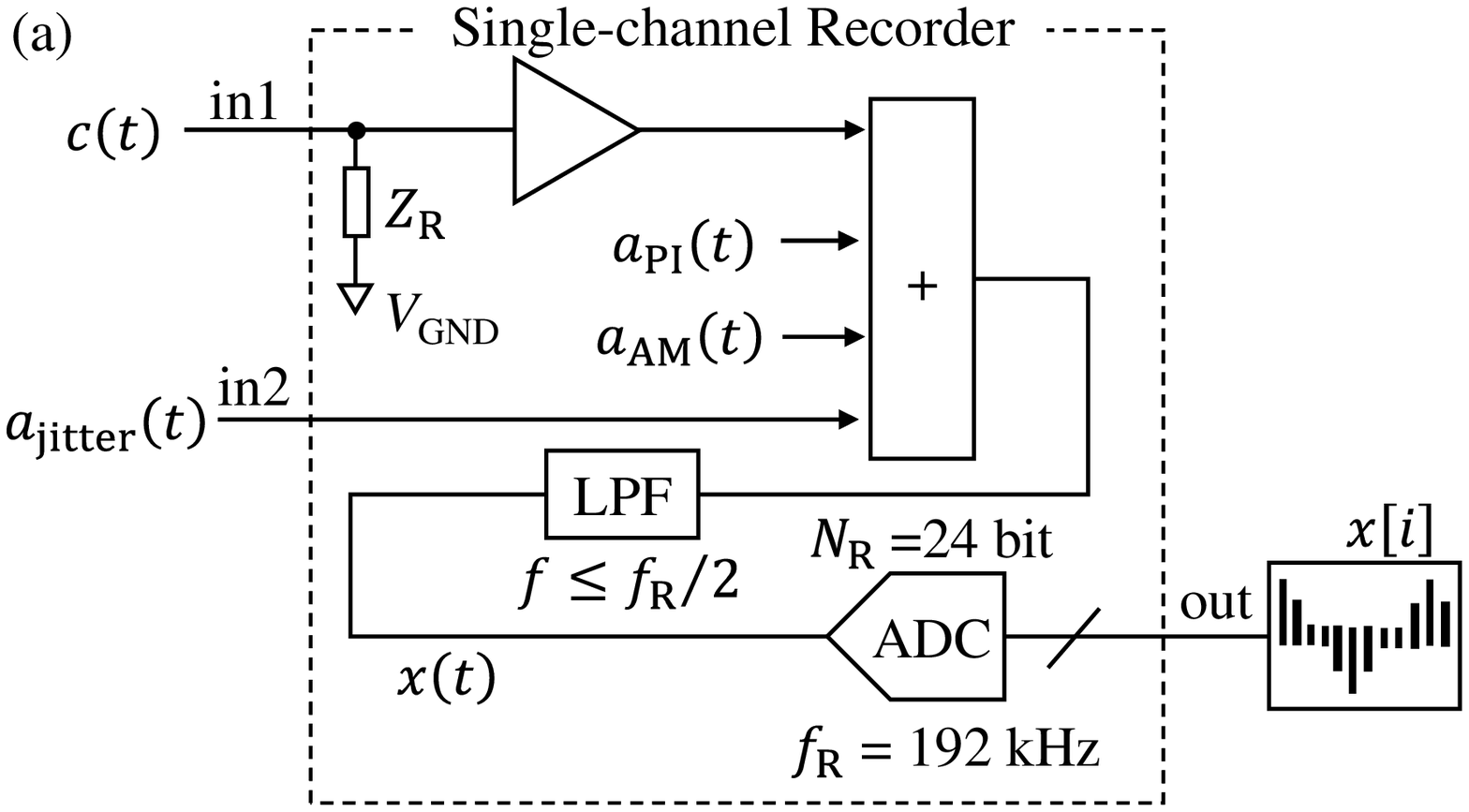}{0.45\figwidth}{}
\fig{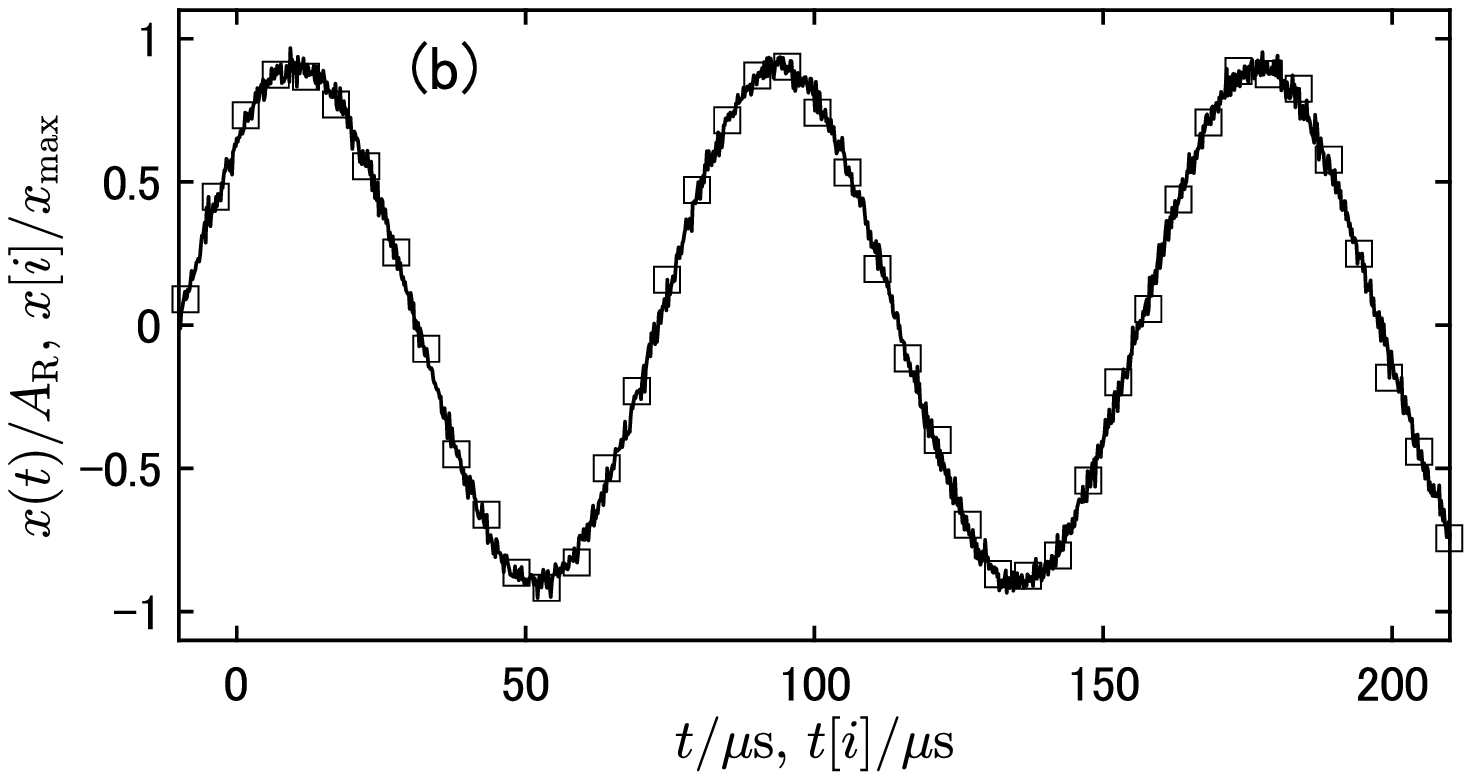}{0.45\figwidth}{}
\fig{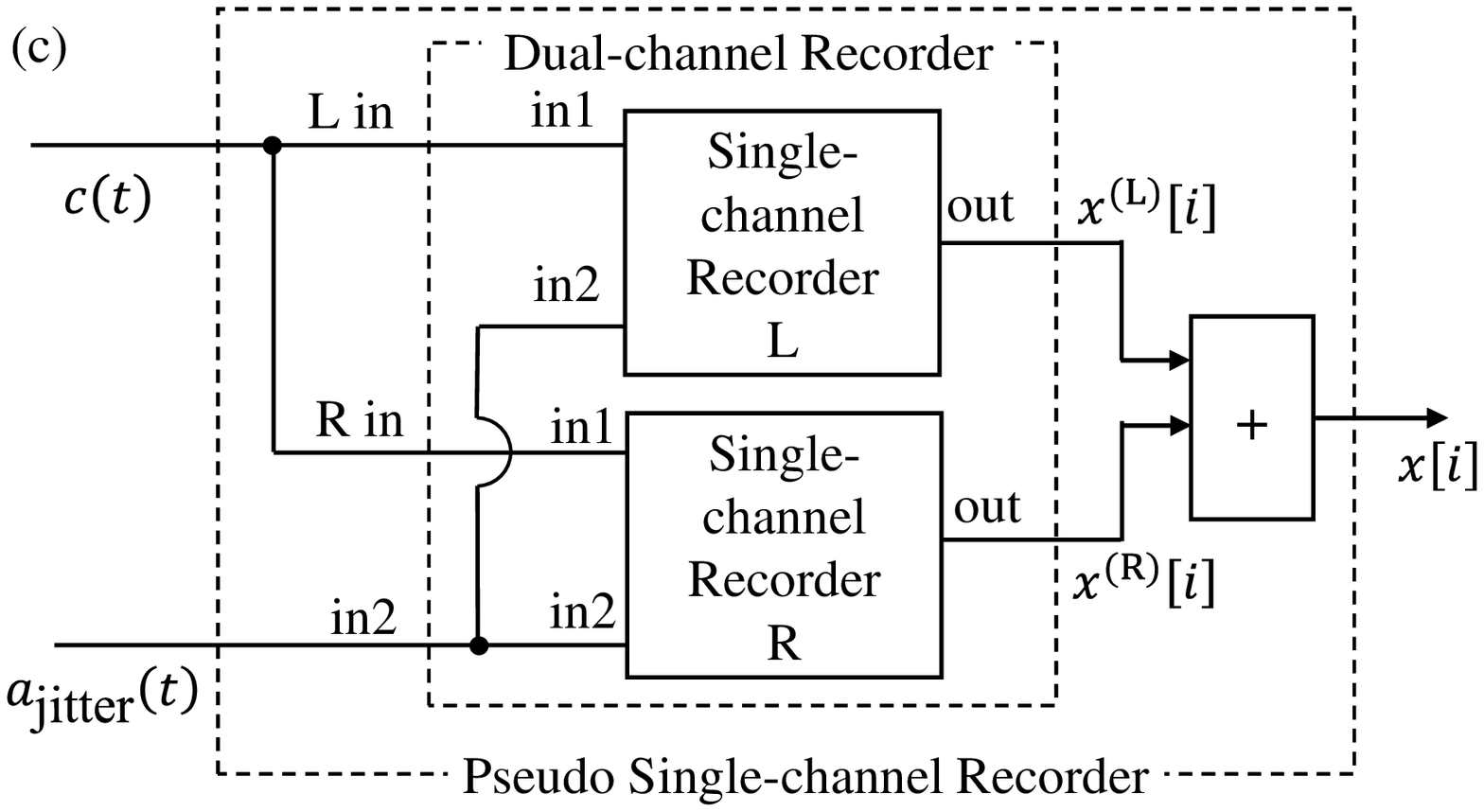}{0.45\figwidth}{}}
\caption{\label{Fig:recorder}{(a) Model diagram of a single-channel digital audio recorder. (b) Relationship between signal before LPF $x(t)$ (solid line) and recorded waveform $x_i$ (white square). (c) Model diagram of a dual-channel digital audio recorder. The method to utilize a dual-channel recorder as a single-channel recorder is depicted.}}
\end{figure}

To represent jitter, AM, and PI noises in a real recording process, we present a model diagram of a single-channel digital audio recorder in \autoref{Fig:recorder}(a). 
As in the previous section, the parameters are fixed to be the same as the experimental conditions described in Section \ref{Section4}. The playback signals $c(t)$ are fed into a buffer amplifier with input impedance $Z_\mathrm{R}$. Subsequently, jitter $a_\mathrm{jitter}(t)$, AM $a_\mathrm{AM}(t)$, and PI noise $a_\mathrm{PI}(t)$ are added, and the high-frequency component is attenuated by an LPF. The cut-off frequency of the LPF is assumed to be $f_\mathrm{R}/2$. The voltage signal before an ADC is represented as $x(t)$. Their relationship is similar to Eqs. (\ref{e6}) and (\ref{Eq:c}). This relationship can thus be expressed as follows: 
\linenomath
\begin{align}
a_\mathrm{total}(t)&=a_\mathrm{jitter}(t)+a_\mathrm{AM}(t)+a_\mathrm{PI}(t), \label{Eq:a_total}\\
x(t)&=\mathcal{LF}\{c(t)+a_\mathrm{total}(t)\},
\end{align}
where $\mathcal{LF}\{~\}$ denotes the low-frequency component. 

The ADC in the recorder, assumed to be ideal and noise-free, performs conversion at $N_\mathrm{R}$ bits with a sampling rate of $f_\mathrm{R}$. The recorded waveform is represented as $x[i]$. The $i$th value is expressed as $x[i]=\mathrm{floor}[x_\mathrm{max}\{x(t[i])/A_\mathrm{R}\}]$, where $A_\mathrm{R}$ is a constant with voltage dimensions, and $x_\mathrm{max}:=2^{N_\mathrm{R}-1}-1$ is the maximum value of a $N_\mathrm{R}$ bit signed integer. In this study, $N_\mathrm{R}$ and $f_\mathrm{R}$ are set to $24~\mathrm{bit}$ and $192~\mathrm{kHz}$, respectively. The analog-to-digital conversion timing is denoted as $t[i]$. The value $t[i]$ is expressed as 
\linenomath
\begin{align}
t[i] &= t_\mathrm{R}+(i-1)f_\mathrm{R}^{-1}, \label{Eq:f_R}
\end{align}
where $t_\mathrm{R}$ represents the time at which the recording started. We assume that the ADC begins working before the playback starts and that the length of $t[i]$ is adequate to record the entire playback signal. When the voltage signal before the ADC is $x(t[i])=A_\mathrm{R}$, the recorded waveform becomes $x[i]=x_\mathrm{max}$. In the real audio recorder, jitter results from the fluctuation of $f_\mathrm{R}$; however, in this model, the LPF and ADC are ideal, and jitter is added before the LPF.

The relationship between $x(t)$ and $x[i]$ is depicted in \autoref{Fig:recorder}(b). The solid line represents $x(t)$. The recorded waveform $x[i]$ is plotted as white squares at $t=t[i]$ and $x(t)=x[i]$. The range of the horizontal axis is equal to that of \autoref{Fig:player}(c). The fluctuation of the solid line in this figure shows artificial random noise. Because the ratio between the sampling rate of the ADC and the frequency $F_\mathrm{pure}(t)$ is $f_\mathrm{R}/f_\mathrm{C}=16$, 16 sampling points are present for each wavelength\footnote{The player yields a sinusoidal wave of $f_\mathrm{C}=f_\mathrm{P}/4$; however, the frequency measured by the recorder is not exactly $f_\mathrm{P}/4$.}.

We demonstrate the model diagram of a dual-channel digital audio recorder in \autoref{Fig:recorder}(c). The recorder comprises two single-channel digital audio recorders; the jitter inputs of the single-channel recorders are assumed to be equipotential. To obtain the experimental results described in Sections \ref{Sec:result_single}, \ref{Sec:result_double}, and \ref{Sec:result_njitter}, we used a dual-channel recorder as a single-channel recorder by contacting two analog inputs and averaging their two waveforms $x^{(L)}[i]$ and $x^{(R)}[i]$. We term this setup a ``pseudo single-channel recorder.'' As an exception, we analyzed waveforms $x^{(L)}[i]$ and $x^{(R)}[i]$ separately to estimate the jitter from the digital audio recorder. See Section \ref{Sec:result_ajitter} for details.

\subsection{\label{Sec:ZCA}ZCA}
In ZCA, we first seek the time at which the voltage signal in the recorder, i.e., $x(t)$, crosses the $t$-axis while $0 \leqslant t \leqslant T$. For this purpose, we reconstruct a continuous function $x'(t)$ from sampling data $x[i]$, which satisfies $x'(t)\approx x(t)$ for $0 \leqslant t \leqslant T$. The reconstruction process comprises three steps. (i) To avoid the boundary effect of the sampling data, a window function $w(t[i])$ is multiplied to $x[i]$ as $x[i]w(t[i])$, where $w(t)$ is the Blackman type, and is expressed as follows: 
\linenomath
\begin{align}
w(t)=
\left\{
\begin{array}{lr}
\multicolumn{2}{l}{0.42+0.5 \cos (\pi f_\mathrm{R} t/N)+0.08 \cos (2\pi f_\mathrm{R} t/N)}\\
\multicolumn{2}{r}{(-Nf_\mathrm{R}^{-1}\leqslant t<0)} \\
~1 & (0 \leqslant t \leqslant T) \\
w(T-t) & (T < t \leqslant T+Nf_\mathrm{R}^{-1}).\\
\end{array}
\right.
\end{align}
We set $N=48~000$ and $T=4 N f_\mathrm{R}^{-1}\approx 1~\mathrm{s}$. Consequently, the data length $w(t[i])$ becomes $6N$, and the domain of $w(t)$ becomes $-0.25~\mathrm{s} \leqslant t \leqslant 1.25~\mathrm{s}$. Thus the data length $x[i]w(t[i])$ becomes $6N$. (ii) After the multiplication with window function, the data points are interpolated using the fast Fourier transform (FFT) method by an oversampling factor of $N_\mathrm{over}=64$. As a result, the number of data points increases to $6N_\mathrm{over}N$. The value of $N_\mathrm{over}$ is adjusted depending on the required accuracy. We confirmed that $N_\mathrm{over}=64$ is sufficiently large by performing numerical simulation\footnote{See Appendix for the results of numerical simulation}. Thanks to the FFTW library used in MATLAB\citep{FFTWsite}, the computation time is almost negligible. We also applied bandwidth limitations to the data to eliminate the DC component. (iii) The interpolated points are connected by a line. After these three steps, a continuous function $x'(t)$ is obtained from the discrete data $\{x[i_\mathrm{main}-N],\cdots, x[i_\mathrm{main}+5N]\}$.

\begin{figure}
\figcolumn{\fig{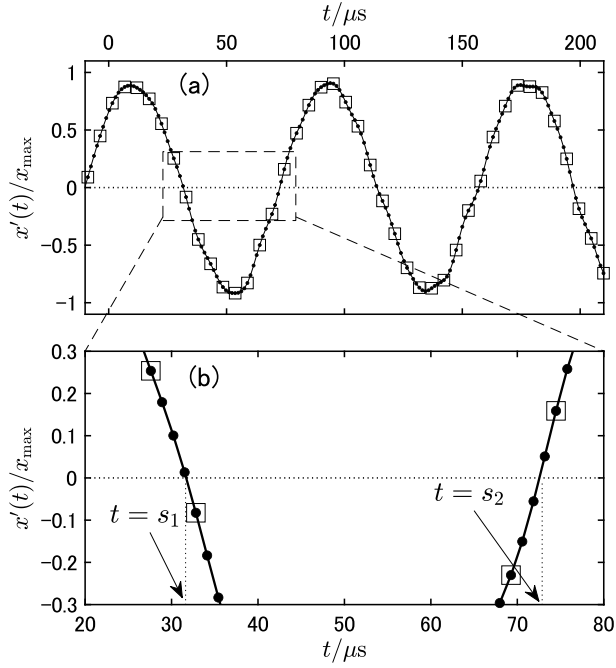}{0.45\figwidth}{}}
\caption{\label{Fig:zca}{(a) The recorded waveform $x_i$ (white square), points obtained by FFT interpolation (black circle), and the continuous function $x'(t)$ (solid line). For ease of viewing, the FFT interpolation was performed using an oversampling factor of $N_\mathrm{over}=4$ rather than $N_\mathrm{over}=64$. (b) Magnified graph for $20~\rmu\mathrm{s} \leqslant t\leqslant 80~\rmu\mathrm{s}$. }}
\end{figure}

\autoref{Fig:zca}(a) shows the process to obtain $x'(t)$. 
The range of the horizontal axis is equal to that of \autoref{Fig:recorder}(b). The white squares are the recorded waveform $x[i]$, and the black circles are the interpolated points by the FFT method. Solid lines represent $x'(t)$. \autoref{Fig:zca}(b) presents a magnified view of \autoref{Fig:zca}(a) around the first and second ZCPs. For the ease of viewing, the FFT interpolation was performed using an oversampling factor $N_\mathrm{over}=4$ in \autoref{Fig:zca}(a)(b). One can see that the $N_\mathrm{over}-1$ points are interpolated between two recorded data points. As shown in \autoref{Fig:zca}(b), the zero-crossing times are labeled as $t=s_1, s_2, \cdots, s_M$, where $M$ denotes the number of ZCPs when $0\leqslant t\leqslant T$. The obtained sequence $s_1, s_2, \cdots, s_M$ is not equally spaced because of jitter. 

\begin{figure}
\figcolumn{\fig{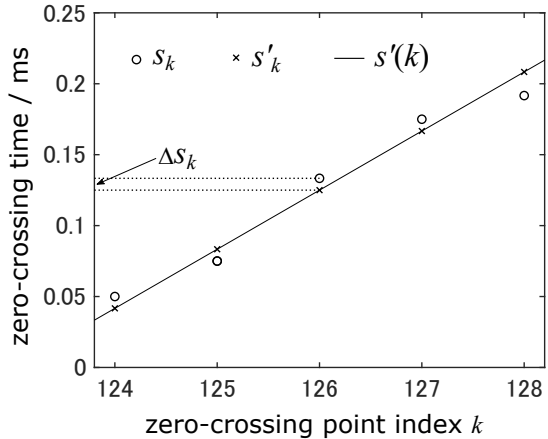}{0.4\figwidth}{}}
\caption{\label{Figure3}{The zero-crossing time in ms versus zero-crossing index $k$. $s_k$ denotes the zero-crossing time of the recorded data, and $s'_k$ denotes that of the corresponding pure sinusoidal wave. Consecutive times $s'_k$ are equally spaced, whereas $s_k$ are not. The difference between $s_k$ and $s'_k$ is extremely magnified for the ease of viewing.}}
\end{figure}

Second, we identify equally spaced points $s'_k$, which are introduced in Eq. (\ref{Eq:s'_k}) in ZCA. For this purpose, a straight line is fitted to $s_k$ using the least-squares method. A conceptual diagram is provided in \autoref{Figure3}. 
The fitting function is written as follows: 
\linenomath
\begin{align}
s'(k)=\frac{k-1}{2f'_\mathrm{C}}+s'_1, \label{Eq:straight}
\end{align}
where $f'_\mathrm{C}$ is the frequency of the playback signal measured by the recorder. Deviation in $s_k$ from the straight line is less than $100~\mathrm{ps}$; this is enlarged in \autoref{Figure3} to ease visualization. From the fitting function of Eq. (\ref{Eq:straight}), the $k$th equidistant point $s'_k:=s'(k)$ is obtained. The frequency $f'_\mathrm{C}$ is the averaged frequency during $0\leqslant t\leqslant T$. Thus, this analysis is sensitive to short-term drift with a frequency of $f\geqslant 1/T$ but is not sensitive to long-term drift. 

In ZCA, we finally obtain the ZCF that gives the difference between $s_k$ and $s'_k$, which is expressed as follows: 
\linenomath
\begin{align}
\varDelta s_k=s'_k-s_k. 
\end{align}
ZCF $\varDelta s_k$ goes to jitter $j(s'_k)$ when both PI noise $n_\mathrm{PI}(t)$ and recorder noise $a_\mathrm{total}(t)$ are negligible. Because the reconstructed function $x'(t)$ crosses the $t$-axis twice per cycle, one can obtain ZCF values at a repetition rate of $f_\mathrm{Z}=2f_\mathrm{C}$. In other words, the bandwidth of $j(t)$ reconstructed from $\varDelta s_k$ becomes $f\leqslant f_\mathrm{Z}/2=f_\mathrm{C}$. This is expected because jitter resembles the frequency modulation in which it is impossible to transmit a frequency higher than the carrier wave. If one observes only the rising or falling ZCPs, the bandwidth of $j(t)$ becomes restricted to $f\leqslant f_\mathrm{C}/2$, which is insufficient to perfectly reconstruct $j(t)$.

\subsection{\label{Sec:double}Single recorder setup (SRS) and DRS}
\begin{figure}
\figcolumn{
\fig{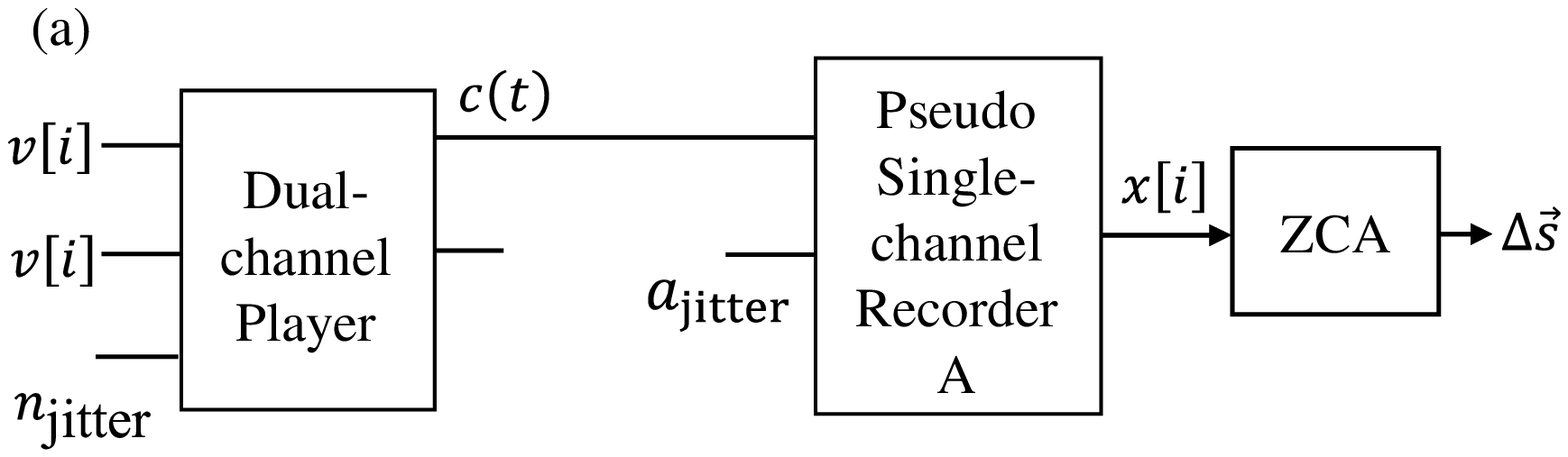}{0.45\figwidth}{}
\fig{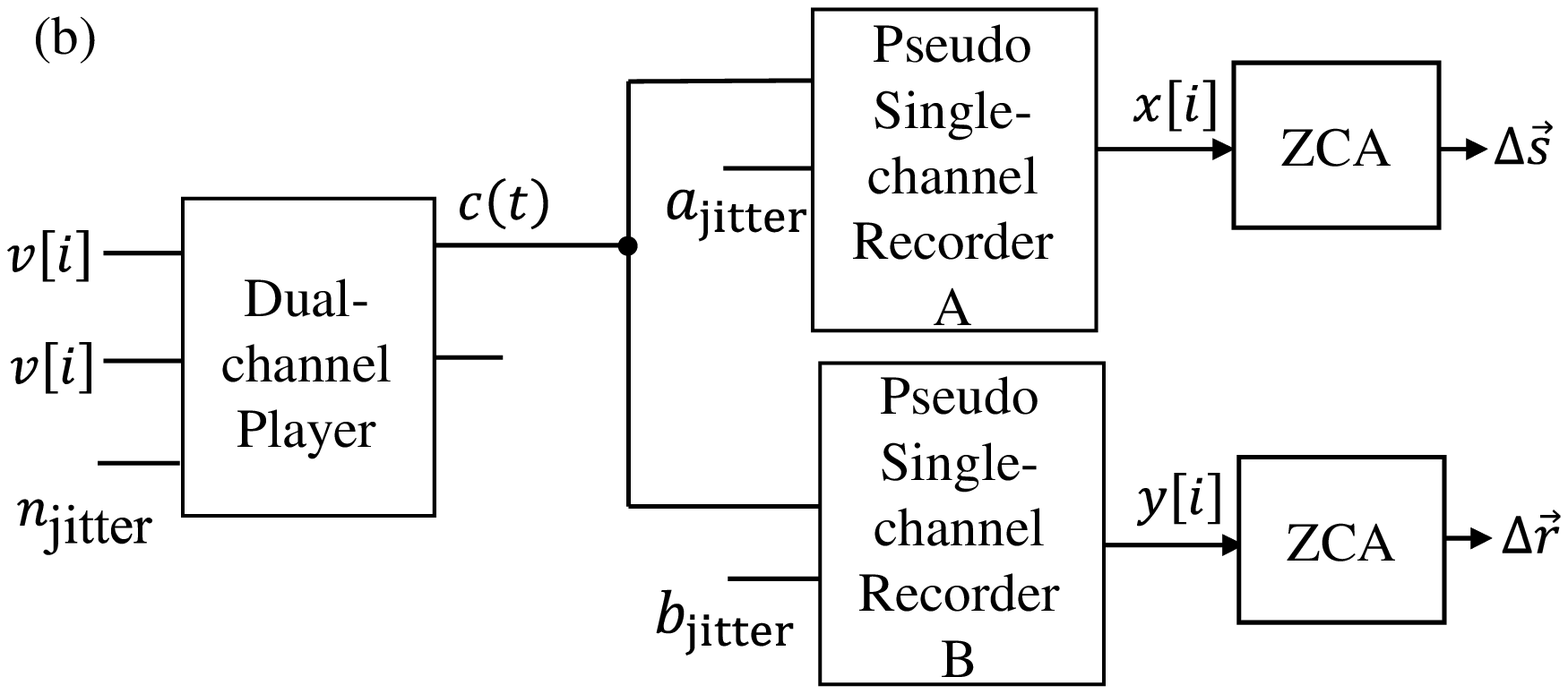}{0.45\figwidth}{}
\fig{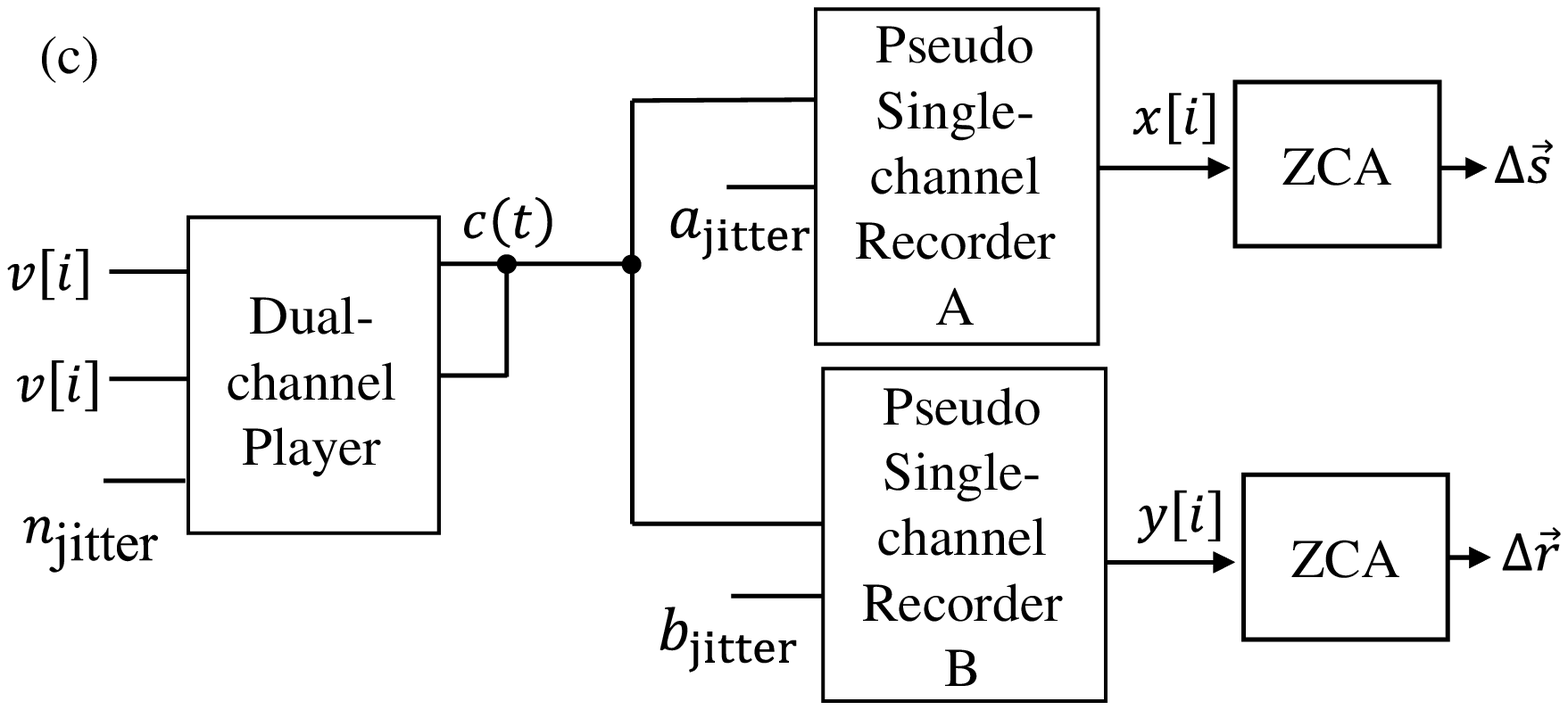}{0.45\figwidth}{}}
\caption{\label{Fig:double}{(a) SRS. (b) DRS. (c) Setup for separating jitter from PI noise.}}
\end{figure}

First, we consider the case in which one player and one recorder are connected [\autoref{Fig:double}(a)]. 
We term this setup a ``single recorder setup (SRS).'' In an SRS, both the player and recorder noises are included in ZCFs and are represented as $\varDelta s_k$. The relationships amongst the player noise, recorder noise, and ZCFs are expressed as follows: 
\linenomath
\begin{align}
(-1)^k\omega A_0\varDelta s_{k}=
&n_\mathrm{jitter}(s'_k)+n_\mathrm{PI}(s'_k) \nonumber\\
&\quad +a_\mathrm{jitter}(s'_k)+a_\mathrm{PI}(s'_k). \label{Eq:single_ZCF}
\end{align}
In the following, $\mathbb{V}\{~\}$ denotes the variance of data. From Eq. (\ref{Eq:single_ZCF}), the variance of ZCFs $\mathbb{V}\{\varDelta s_k\}$ becomes
\linenomath
\begin{align}
\mathbb{V}\{\varDelta s_k\}&=(\sigma_{n1})^2+(\sigma_{a1})^2, \label{Eq:single}
\end{align}
where $\sigma_{n1}$ and $\sigma_{a1}$ are the root mean squares (RMSs) of ZCFs for the player and the recorder, respectively. More explicitly, $\sigma_{n1}$ and $\sigma_{a1}$ can be expressed as follows: 
\linenomath
\begin{align}
(\sigma_{n1})^2&=\mathbb{V}\{j(s'_k)\}+\frac{\mathbb{V}\{n_\mathrm{PI}(s'_k)\}}{(\omega A_0)^2}, \label{Eq:sigma_n}\\ 
(\sigma_{a1})^2&=\frac{\mathbb{V}\{a_\mathrm{jitter}(s'_k)\}+\mathbb{V}\{a_\mathrm{PI}(s'_k)\}}{(\omega A_0)^2}.
\end{align}
The left-hand side of Eq. (\ref{Eq:single}) can be obtained using experimental data, the results of which are described in Section \ref{Sec:result_single}. 

Second, we consider the case in which one player and two recorders are connected [\autoref{Fig:double}(b)]; this setup is termed a ``DRS.'' In the DRS, both player and recorder noises are included in each ZCF, which are represented as $\varDelta s_k$ and $\varDelta r_k$. The relation of Eq. (\ref{Eq:single_ZCF}) holds for the DRS. As in Eq. (\ref{Eq:single_ZCF}), $\varDelta r_k$ is expressed as 
\linenomath
\begin{align}
(-1)^k\omega A_0\varDelta r_{k}=
&n_\mathrm{jitter}(r'_k)+n_\mathrm{PI}(r'_k) \nonumber\\
&\quad +b_\mathrm{jitter}(r'_k)+b_\mathrm{PI}(r'_k), \label{Eq:b_ZCF}
\end{align}
where $r'_k$ is the equally spaced time obtained by ZCA using a waveform acquired by recorder B (represented as $y[i]$), $b_\mathrm{jitter}(t)$ is the jitter of recorder B, and $b_\mathrm{PI}(t)$ is the PI noise of recorder B. Because the two recorders simultaneously sample the same playback signal outputted from one player, we can say $s'_k=r'_k$ in the real world. Note that $s'_k$ and $r'_k$ have a common index $k$. Consequently, $r'_k$ in Eq. (\ref{Eq:b_ZCF}) can be replaced by $s'_k$ then, we obtain the following: 
\linenomath
\begin{align}
(-1)^k\omega A_0\varDelta r_{k}=
&n_\mathrm{jitter}(s'_k)+n_\mathrm{PI}(s'_k) \nonumber\\
&\quad +b_\mathrm{jitter}(s'_k)+b_\mathrm{PI}(s'_k). \label{Eq:b_ZCF_s'}
\end{align}
Using Eqs. (\ref{Eq:single_ZCF}) and (\ref{Eq:b_ZCF_s'}), the following four equations are obtained: 
\linenomath
\begin{align}
\mathbb{V}\{\varDelta s_k\}&=(\sigma_{n2})^2+(\sigma_{a2})^2, \label{e22-3}\\
\mathbb{V}\{\varDelta r_k\}&=(\sigma_{n2})^2+(\sigma_{b2})^2,\label{e22-4}\\
\mathbb{V}\{\varDelta s_k-\varDelta r_k\}&=(\sigma_{a2})^2+(\sigma_{b2})^2, \label{e22-5}\\
\mathbb{V}\{\varDelta s_k+\varDelta r_k\}&=4(\sigma_{n2})^2+(\sigma_{a2})^2+(\sigma_{b2})^2, \label{e22-6}
\end{align}
where $\sigma_{n2}$, $\sigma_{a2}$, and $\sigma_{b2}$ are the RMSs of ZCFs for the player, recorder A, and recorder B, respectively. It is important to note that the effect of the player is canceled in $\varDelta s_k-\varDelta r_k$. Moreover, the player makes a double contribution in $\varDelta s_k+\varDelta r_k$. This is because $s'_k$ and $r'_k$ are common to both instruments, even though $\varDelta s_k$ and $\varDelta r_k$ are measured on different instruments. 

The left-hand side of these equations can be obtained using experimental data. Using Eqs. (\ref{e22-3})--(\ref{e22-5}), we can evaluate noise in the player ($\sigma_{n2}$) separately from that in the recorders ($\sigma_{a2}$ and $\sigma_{b2}$). Eq. (\ref{e22-6}) can be used to verify the calculations. The experimental results are presented in Subsection \ref{Sec:result_double}. 

These noises, $\sigma_{n2}$,  $\sigma_{a2}$, and $\sigma_{b2}$, can be expressed as 
\linenomath
\begin{align}
(\sigma_{n2})^2&=\mathbb{V}\{j(s'_k)\}+\frac{\mathbb{V}\{n_\mathrm{PI}(s'_k)\}}{(\omega A_0)^2}, \label{Eq:sigma_n2}\\ 
(\sigma_{a2})^2&=\frac{\mathbb{V}\{a_\mathrm{jitter}(s'_k)\}+\mathbb{V}\{a_\mathrm{PI}(s'_k)\}}{(\omega A_0)^2}, \label{Eq:sigma_a2}\\
(\sigma_{b2})^2&=\frac{\mathbb{V}\{b_\mathrm{jitter}(s'_k)\}+\mathbb{V}\{b_\mathrm{PI}(s'_k)\}}{(\omega A_0)^2}. \label{Eq:sigma_b2}
\end{align}

\subsection{\label{Sec:jitter-PI}Separating jitter from PI noise}
Finally, we consider the case shown in \autoref{Fig:double}(c). This setup enables us to separate jitter from PI noise for the player. The setup differs from that of \autoref{Fig:double}(b),where the L and R signals of the player are bundled together. The RMS of ZCFs for the player, represented as $\sigma_{n3}$, can be obtained as in the DRS. 

As shown in \autoref{Fig:player}(d), the dual-channel player comprises two single-channel players with equivalent jitter. Moreover, the PI noises of the two single-channel players are independent. Therefore, under the assumptions of this model, PI noise in $\sigma_{n3}$ is reduced to
\linenomath
\begin{align}
(\sigma_{n3})^2 &=\mathbb{V}\{j(s'_k)\}+\frac{1}{2}\frac{\mathbb{V}\{n_\mathrm{PI}(s'_k)\}}{(\omega A_0)^2}. \label{Eq:sigma_n3}
\end{align}
Using Eqs. (\ref{Eq:sigma_n2}) and (\ref{Eq:sigma_n3}), we can obtain $\mathbb{V}\{j(s'_k)\}$. Therefore, the RMS of jitter can be determined by measuring $\sigma_{n2}$ and $\sigma_{n3}$. This relationship is expressed as 
\linenomath
\begin{align}
\mathrm{dev}\{j(s'_k)\}=\sqrt{2(\sigma_{n3})^2-(\sigma_{n2})^2},
\end{align}
where $\mathrm{dev}\{~\}$ denotes the deviation, i.e., $\mathrm{dev}\{~\}:=\sqrt{\mathbb{V}\{~\}}$. 

\section{\label{Section4}Experimental procedure}
\subsection{Audio players and recorders}
In our experiment, we used three identical portable audio devices (DR-100MKIII; TASCAM, Japan). These devices offer several advantages: they are unaffected by the quality of the alternating current power supply, which ensures the reproducibility and independency of the measurement, are inexpensive, and are easy to obtain. 

\begin{figure}
\figcolumn{\fig{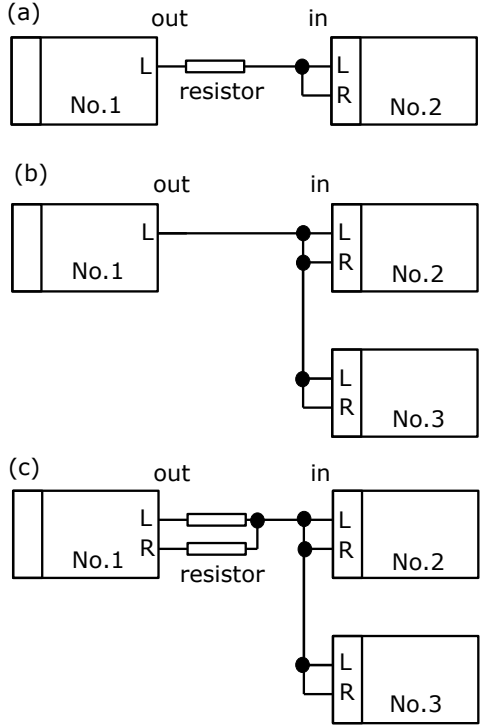}{0.35\figwidth}{}}
\caption{\label{Figure12}{(a) SRS; The output, the left channel of device No. 1, was fed simultaneously to the left and right channels of device No. 2. A matching resistor of $200~\mathrm{\Omega}$ was inserted. (b) DRS; The output, the left channel of device No. 1, was fed simultaneously to the left and right channels of device No. 2 and No. 3. (c) The output is the sum of the left and right channels of device No. 1, and is fed simultaneously to the left and right channels of device No. 2 and No. 3.}}
\end{figure}

One of the three devices (No. 1) was used as a player, and the others (No. 2 and No. 3) were used as recorders. \autoref{Figure12}(a) and (b), corresponds to SRS and DRS, respectively, as described in Subsection \ref{Sec:double}. 
\autoref{Figure12}(c) shows the setup required to separate jitter from PI noise as described in Subsection \ref{Sec:jitter-PI}. 

The device settings for the recorders are summarized in Table \ref{Table1}. 
\begin{table}[ht]
\caption{\label{Table1}Device settings for the recorders.}
\begin{tabular}{clll}
\hline
FILE FORMAT & WAV24\\
SAMPLING RATE & 192kHz\\
FILE TYPE & STEREO\\
XRI & OFF\\
DUAL REC & OFF\\
\hline
SOURCE & EXT LINE\\
A/D FILTER & FIR1\\
DUAL ADC & ON\\
LOW CUT & OFF\\
\hline
RECORDING LEVEL & +3dB\\
\hline
\end{tabular}
\end{table}
The recording levels for the three setups are adjusted to be equal by inserting a matching resistor as shown in \autoref{Figure12}(a)(c). This is necessary to prevent level changes in recordings that affect PI noise.

\subsection{How to synchronize different waveforms}
Details of the playback file are described in Subsection \ref{Sec:player}. The length of the main part is $N_\mathrm{main}f_\mathrm{P}^{-1}\approx 30~\mathrm{s}$. The lengths of the fade-in and fade-out parts are both $N_\mathrm{F}f_\mathrm{P}^{-1}\approx 5~\mathrm{s}$. Therefore, there are $(N_\mathrm{main}+2N_\mathrm{F})/4=480~000~\mathrm{cycles}$ of sinusoidal waves in the playback signal, and the same is true for the recorded waveform. In our improved TDA, the analysis program counts the number of cycles in the two sinusoidal waves from two recorders and assigns a common zero-crossing index. This characteristic is of importance in Subsection \ref{Sec:result_double}.

\section{\label{Sec:result_all}Results of measurements}
\subsection{\label{Sec:result_single}SRS}
\begin{figure}
\figcolumn{\fig{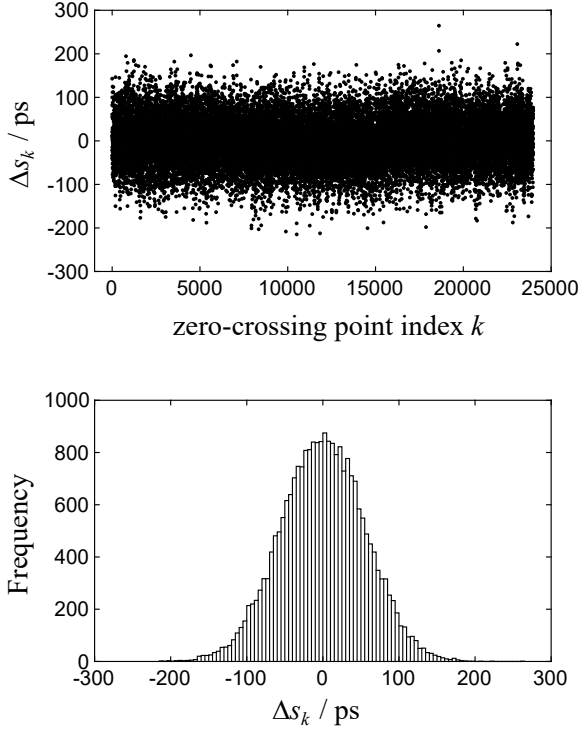}{0.425\figwidth}{}}
\caption{\label{Figure13}{(a) ZCF obtained for 1 s. There are $M=24~000$ ZCPs. (b) Histogram of the obtained ZCF. }}
\end{figure}

Using the SRS [\autoref{Figure12}(a)], we played back and recorded the sinusoidal wave of $f_\mathrm{C}=12~\mathrm{kHz}$. To eliminate low-frequency noise that does not originate from the clock in the player, the recorded waveform was processed in a limited bandwidth range of $f_\mathrm{C}-B_\mathrm{w}\leqslant f\leqslant f_\mathrm{C}+B_\mathrm{w}$. Consequently, we analyzed jitter in the bandwidth of $1/T\leqslant f\leqslant B_\mathrm{w}$. In this analysis, $B_\mathrm{W}$ was set to $6~\mathrm{kHz}$. Then, the ZCF was obtained using MATLAB code. Figure \ref{Figure13}(a) shows the obtained ZCF, $\varDelta s_1$, $\varDelta s_2$, $\cdots$, and $\varDelta s_M$. 
Figure \ref{Figure13}(b) shows the distribution of the obtained ZCF, which resembles a Gaussian curve. As expressed in Eq. (\ref{Eq:single}), this ZCF includes the effects of jitter and PI noise from both the player and recorder. The RMS of ZCF is $\{(\sigma_{n1})^2+(\sigma_{a1})^2\}^{1/2}=55.3~\mathrm{ps}$. 

\subsection{\label{Sec:result_double}DRS}
\begin{figure}
\figcolumn{\fig{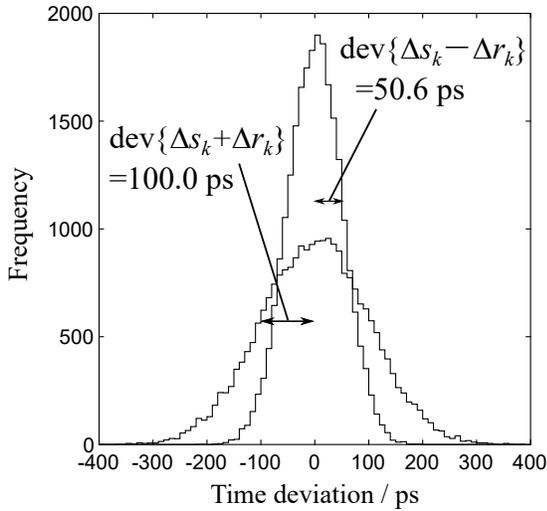}{0.4\figwidth}{}}
\caption{\label{Figure15}{Histograms of $\varDelta s_k-\varDelta r_k$ and $\varDelta s_k+\varDelta r_k$. The RMS values of the former and latter were $50.6$ and $100.0~\mathrm{ps}$, respectively.}}
\end{figure}

Using the DRS [\autoref{Figure12}(b)], we obtained the ZCFs $\varDelta s_k$ and $\varDelta r_k$ from devices No. 2 and No. 3, respectively. \autoref{Figure15} shows the distributions of $\varDelta s_k-\varDelta r_k$ and $\varDelta s_k+\varDelta r_k$, wherein the former is clearly narrower than the latter. 
This is because the effect of the player is canceled out in $\varDelta s_k-\varDelta r_k$, whereas it makes a double contribution in $\varDelta s_k+\varDelta r_k$. From the experimental data, we can determine
\linenomath
\begin{align}
E_1
&=\mathrm{dev}\{\varDelta s_k\},\\
E_2
&=\mathrm{dev}\{\varDelta r_k\},\\
E_3
&=\mathrm{dev}\{\varDelta s_k-\varDelta r_k\},\\
E_4
&=\mathrm{dev}\{\varDelta s_k+\varDelta r_k\},
\end{align}
where $E_1$, $E_2$, $E_3$, and $E_4$ are the standard deviations calculated from $\{\varDelta s_1,\cdots, \varDelta s_M\}$, $\{\varDelta r_1,\cdots, \varDelta r_M\}$, $\{\varDelta s_1-\varDelta r_1,\cdots, \varDelta s_M-\varDelta r_M\}$, and $\{\varDelta s_1+\varDelta r_1,\cdots, \varDelta s_M+\varDelta r_M\}$, respectively. The values of $E_1$, $E_2$, $E_3$, and $E_4$ are obtained as $E_1=56.0~\mathrm{ps}$, $E_2=56.1~\mathrm{ps}$, $E_3=50.6~\mathrm{ps}$, and $E_4=100.0~\mathrm{ps}$, respectively. Using equations (\ref{e22-3}), (\ref{e22-4}), and (\ref{e22-5}), we obtain the following RMS values of ZCFs for the player: 
\linenomath
\begin{align}
\sigma_{n2} &=43.1~\mathrm{ps}, \label{e422-5}
\end{align}
and those of the recorders as 
\linenomath
\begin{align}
\sigma_{a2} &=35.7~\mathrm{ps}, \label{e422-6}\\
\sigma_{b2}&=35.9~\mathrm{ps}. \label{e422-7}
\end{align}
These values satisfy Eq. (\ref{e22-6}), demonstrating that there is no correlation between the two recorders. 

\subsection{\label{Sec:result_njitter}Jitter and PI noise of player}
In this subsection, we obtain the jitter and PI noise of the player separately. We measured a ZCF using the setup shown in \autoref{Figure12}(c). Using the same method as in the previous subsection, we determined the RMS values of the ZCF of the player, $\sigma_{n3}$. The obtained value is 
\linenomath
\begin{align}
\sigma_{n3}=33.5~\mathrm{ps}.
\end{align} 
This is smaller than $\sigma_{n2}=43.1~\mathrm{ps}$ calculated in the previous subsection. As expressed by Eqs. (\ref{Eq:sigma_n2})--(\ref{Eq:sigma_n3}), this difference can be interpreted as a decrease in PI noise due to averaging. From these, we obtain
\linenomath
\begin{align}
\mathrm{dev}\{j(s'_k)\} &=19.7~\mathrm{ps}, \label{Eq:dev_j}\\
\frac{\mathrm{dev}\{n_\mathrm{PI}(s'_k)\}}{\omega A_0}&=38.4~\mathrm{ps}.  
\end{align}
These results indicate that PI noise must be considered when evaluating the jitter in recent audio equipment with small jitter values, typically less than 100~ps. As noted in Section \ref{Sec:result_single}, jitter was calculated in the bandwidth of $1~\mathrm{Hz}\leqslant f\leqslant 6~\mathrm{kHz}$. Meanwhile, PI noise was analyzed in the bandwidth of $6~\mathrm{kHz}\leqslant f\leqslant 18~\mathrm{kHz}$.

\section{\label{Section5} Discussions}
\subsection{Detection limit of jitter}
Using the proposed method, jitter can be measured with higher accuracy than when using existing methods\citep{Dunn92,Dunn94Feb,Dunn94May,Dunn00,Nishimura10}. This is partly due to the recent improvement in the performance of ADCs. We measured waveforms at 192~kHz and 24~bit, whereas 16-bit DACs of 44.1 or 48~kHz were used in previous studies\citep{Nishimura10}. The detection limit of the proposed method exists due to quantization noise, which is represented as $j_\mathrm{LSB}$ and can be obtained by solving the following equation: 
\linenomath
\begin{align}
x_\mathrm{max}\frac{A_0\omega j_\mathrm{LSB}}{A_\mathrm{R}}=1.
\end{align}
The result becomes $j_\mathrm{LSB} \approx 1.76~\mathrm{ps}$.

The mean for $\sigma_{n2}$ and that for $\mathrm{dev}\{j(s'_k)\}$ are obtained as follows: 
\linenomath
\begin{align}
\sigma_{n2} &=42.34 (14)~\mathrm{ps},\\
\mathrm{dev}\{j(s'_k)\} &=21.1 (6)~\mathrm{ps},
\end{align}
using the values of $\sigma_{n2}$ and $\mathrm{dev}\{j(s'_k)\} $ for different ten time domains. The numbers in (~) represent the standard deviation of the mean. Therefore, the detection limit lies between $j_\mathrm{LSB}$ and $21.1~\mathrm{ps}$ and is expected to be less than $10~\mathrm{ps}$. Furthermore, the detection limit depends on the recorders employed; more accurate measurements are possible when higher-performance recorders are used.

\subsection{\label{Sec:phase_analysis}Phase dependence of playback noise}
We now consider the phase dependence of the total playback noise, i.e., $n_\mathrm{total}(t)$. One might expect that phase dependence analysis enables the separation of jitter, AM, and PI noise; unfortunately, this approach is not promising, as shown below.

First, we demonstrate that the phase dependence of $n_\mathrm{total}(t)$ can always be expressed by two parameters, $A$ and $B$. For this purpose, we express time $t$ with phase $\theta$ and the number of cycles $m$ as follows: 
\linenomath
\begin{align}
t(\theta,m)=\frac{\theta-\theta_0+2\pi (m-1)}{\omega},
\end{align}
where the maximum value of $m$, represented as $m_\mathrm{max}$, is set to 
\linenomath
\begin{align}
m_\mathrm{max}=\mathrm{floor}\left(\omega T/2\pi \right), \label{Eq:m_max}
\end{align}
and the domain of $\theta$ is restricted to
\linenomath
\begin{align}
0\leqslant \theta < 2\pi. \label{Eq:theta}
\end{align}
Notably, $t(\pi/2,m)=s'_{2 m-1}$ and $t(3\pi/2,m)=s'_{2 m}$. In the following, for simplicity, we replace the expression of $n_\mathrm{total}(t(\theta,m))$ with $n_\mathrm{total}(\theta,m)$. The same rule is also applied to $j(t)$, $A_\mathrm{M}(t)$ and $n_\mathrm{PI}(t)$. As a result, the playback noise is expressed as
\linenomath
\begin{align}
&n_\mathrm{total}(\theta,m) \nonumber\\
&=A_0\omega j(\theta,m)\sin\theta+A_\mathrm{M}(\theta,m)\cos\theta+n_\mathrm{PI}(\theta,m).\label{Eq:n_total}
\end{align}
In the following, $\mathbb{V}\{n_\mathrm{total}(\theta,m)\}$ denotes the variance calculated from $n_\mathrm{total}(\theta,2)$, $n_\mathrm{total}(\theta,3)$, $\cdots$, and $n_\mathrm{total}(\theta,m_\mathrm{max}-1)$. From Eq. (\ref{Eq:n_total}), we obtain
\linenomath
\begin{align}
&\mathbb{V}\{ n_\mathrm{total}(\theta,m)\}
=\frac{1-\cos(2\theta)}{2}(\omega A_0)^2\mathbb{V}\{j(\theta,m)\}\nonumber\\
&\quad\quad+\frac{1+\cos(2\theta)}{2}\mathbb{V}\{A_\mathrm{M}(\theta,m)\}
+\mathbb{V}\{n_\mathrm{PI}(\theta,m)\}\\
&=A\cos(2\theta)+B,
\end{align}
where we set 
\linenomath
\begin{align}
A &:=\frac{\mathbb{V}\{A_\mathrm{M}(\theta,m)\}-(\omega A_0)^2\mathbb{V}\{j(\theta,m)\}}{2},\\
B &:=\mathbb{V}\{n_\mathrm{PI}(\theta,m)\}+\frac{(\omega A_0)^2\mathbb{V}\{j(\theta,m)\}+\mathbb{V}\{A_\mathrm{M}(\theta,m)\}}{2}.
\end{align}
We assume that $\mathbb{V}\{j(\theta,m)\}$, $\mathbb{V}\{A_\mathrm{M}(\theta,m)\}$, and $\mathbb{V}\{n_\mathrm{PI}(\theta,m)\}$ do not depend on $\theta$. Therefore, the phase dependence of $\mathbb{V}\{n_\mathrm{total}(\theta,m)\}$ can always be expressed by two parameters $A$ and $B$ provided the assumptions adopted above are valid.

Consequently, the following behavior can be confirmed: (i) when jitter and AM are not negligible, the offset $B$ is not equal to PI noise; (ii) when jitter and AM are comparable, amplitude $A$ vanishes; (iii) when PI noise is negligible, one can obtain jitter and AM by calculating $B\pm A$; (iv) when PI noise is not negligible, one cannot obtain jitter, AM, and PI noise from $A$ and $B$. Behavior (iv) indicates that further considerations are necessary to separate $\mathbb{V}\{j(\theta,m)\}$ from $B$. The procedure designed for this purpose is explained in Subsections \ref{Sec:jitter-PI} and \ref{Sec:result_njitter}.

\subsection{Comparison with CSM}
As noted in the introduction, the DRS is similar to the setup of CSM\citep{Rubiola10}. CSM can be regarded as a combination of FDA and DRS, whereas the proposed method is a combination of ZCA and DRS. For CSM, noise from two instruments is reduced by averaging, and the cross-spectrum attains the power spectrum of the device under test. 

Commercial products based on CSM are designed to evaluate clock generators with a greater frequency than $1~\mathrm{MHz}$\citep{Feldhaus16Apr}\citep{Feldhaus16AN}. This is primarily because frequency conversion in the audio frequency range is technically challenging. Hence, assessing audio signal with CSM has not been performed so far. The proposed method, a combination of ZCA and DRS, can access audio signal and appears to be feasible as a substitution for CSM.

\subsection{\label{Sec:result_ajitter}Jitter and PI noise of recorder}
In this subsection, we separately obtain the jitter and PI noise of a recorder. As shown in \autoref{Fig:recorder}(c), the dual-channel recorder comprises two single-channel recorders with common jitter. The PI noises of L and R inputs are independent, and are represented as $a_\mathrm{PI,L}(t)$ and $a_\mathrm{PI,R}(t)$, respectively. The ZCFs of the recorded waveforms L and R are represented as $\varDelta s^\mathrm{(L)}_k$ and $\varDelta s^\mathrm{(R)}_k$, respectively. Similar to Eqs. (\ref{e22-3})--(\ref{Eq:sigma_b2}), we obtain the following equations.
\linenomath
\begin{align}
(E_5)^2
&=\mathbb{V}\{\varDelta s^{(\mathrm{L})}_{k}\}\nonumber\\
&=(\sigma_{n2})^2
+\frac{\mathbb{V}\{a_\mathrm{jitter}(s'_{k})\}}{(\omega V_0)^2}
+\frac{\mathbb{V}\{a_\mathrm{PI,L}(s'_{k})\}}{(\omega V_0)^2}, \\
(E_6)^2
&=\mathbb{V}\{\varDelta s^{(\mathrm{R})}_{k}\}\nonumber\\
&=(\sigma_{n2})^2
+\frac{\mathbb{V}\{a_\mathrm{jitter}(s'_{k})\}}{(\omega V_0)^2}
+\frac{\mathbb{V}\{a_\mathrm{PI,R}(s'_{k})\}}{(\omega V_0)^2}, \\
(E_7)^2
&=\mathbb{V}\{\varDelta s^{(\mathrm{L})}_{k}-\varDelta s^{(\mathrm{R})}_{k}\}\nonumber\\
&=\frac{\mathbb{V}\{a_\mathrm{PI,L}(s'_{k})\}+\mathbb{V}\{a_\mathrm{PI,R}(s'_{k})\}}{(\omega V_0)^2}, \\
(E_8)^2
&=\mathbb{V}\{\varDelta s^{(\mathrm{L})}_{k}+\varDelta s^{(\mathrm{R})}_{k}\}\nonumber\\
&=4(\sigma_{n2})^2
+4\frac{\mathbb{V}\{a_\mathrm{jitter}(s'_{k})\}}{(\omega V_0)^2}\nonumber\\
&\quad\quad
+\frac{\mathbb{V}\{a_\mathrm{PI,L}(s'_{k})\}+\mathbb{V}\{a_\mathrm{PI,R}(s'_{k})\}}{(\omega V_0)^2},
\end{align}
where $E_5$, $E_6$, $E_7$, and $E_8$ are the standard deviations calculated from $\{\varDelta s_1^{(L)},\cdots, \varDelta s_M^{(L)}\}$, $\{\varDelta s_1^{(R)},\cdots, \varDelta s_M^{(R)}\}$,  $\{\varDelta s_1^{(L)}-\varDelta s_1^{(R)},\cdots, \varDelta s_M^{(L)}-\varDelta s_M^{(R)}\}$, and $\{\varDelta s_1^{(L)}+\varDelta s_1^{(R)},\cdots, \varDelta s_M^{(L)}+\varDelta s_M^{(R)}\}$, respectively. Using the experimental data generated herein, $E_5$, $E_6$, $E_7$, and $E_8$ are obtained as $ 63.7$, $ 63.1$, $ 61.9$, and $ 110.6~\mathrm{ps}$, respectively. Consequently, we obtain 
\linenomath
\begin{align}
\frac{\mathrm{dev}\{a_\mathrm{PI,L}(s'_{k})\}}{\omega V_0}&=44.3~\mathrm{ps},\\
\frac{\mathrm{dev}\{a_\mathrm{PI,R}(s'_{k})\}}{\omega V_0}&=43.3~\mathrm{ps},
\end{align}
\begin{align}
(\sigma_{n2})^2+\frac{\mathbb{V}\{a_\mathrm{jitter}(s'_{k})\}}{(\omega V_0)^2}&=(45.9 {\rm ps})^2.
\end{align}
With $\sigma_{n2}=43.1~\mathrm{ps}$ in Eq. (\ref{e422-5}), we obtain 
\linenomath
\begin{align}
\frac{\mathrm{dev}\{a_\mathrm{jitter}(s'_{k})\}}{\omega V_0}=15.7~\mathrm{ps}.\label{Eq:dev_ajitter}
\end{align}
The player and recorders used in this experiment are the same product; consequently, we anticipate comparable jitters in the player and the recorders. The results of Eqs. (\ref{Eq:dev_j}) and (\ref{Eq:dev_ajitter}) support this expectation.

\section{\label{Section6}Summary and outlook}
Herein, we proposed an efficient and powerful method for highly accurate jitter measurements. This method is based on two key elements: ZCA and DRS. The ZCA enables us to determine the zero-crossing times of the voltage signals in the recorders ($s_k$ and $r_k$) and those of the pure sinusoidal waves ($s'_k$ and $r'_k$) by analyzing the recorded waveforms ($x[i]$ and $y[i]$). Their respective differences, ``ZCFs ($\varDelta s_k$ and $\varDelta r_k$),'' contain information about both player noise ($\sigma_{n2}$) and recorder noises ($\sigma_{a2}$ and $\sigma_{b2}$). If one measures ZCFs with a DRS, it is possible to eliminate recorder noise from ZCFs by calculating positive and negative correlations between ZCFs ($\mathbb{V}\{\varDelta s_k+\varDelta r_k\}$ and $\mathbb{V}\{\varDelta s_k-\varDelta r_k\}$). As a result, one can independently determine player noise. The player noise ($\sigma_{n2}$) results from the jitter ($\mathrm{dev}\{j(s'_k)\}$) and PI noise ($\mathrm{dev}\{a_\mathrm{PI}(s'_k)\}$). To separate them, some considerations are required. An example of such a procedure is to measure player noise when L and R outputs are bundled together ($\sigma_{n3}$).

We demonstrated the proposed method using commercial audio equipment. The RMS values of jitter and PI noise were determined as $\mathrm{dev}\{j(s'_k)\}\approx 20~\mathrm{ps}$ and $\mathrm{dev}\{n_\mathrm{PI}(s'_k)\}/(\omega A_0)\approx 40~\mathrm{ps}$, respectively. These results show that the proposed method can evaluate values of jitter that are smaller than PI noise. The high accuracy of the proposed method entails that it will be powerful means by which to develop ultrahigh performance devices in the future. Using such devices, more definite and quantitative study of real-life sounds, such as music, becomes possible. This will form the basis of future investigations. 

\appendix*

\section{Numerical simulation}\label{Sec:appendix}
As mentioned in Section \ref{Sec:ZCA}, we developed a ZCA program written in MATLAB code. To confirm that the program can accurately derive ZCFs, we prepared multiple dummies of recorded waveforms in which artificial jitter, AM, PI noise, and recorder noises are added to a pure sinusoidal wave. The dummy waveforms are expressed as follows: 
\linenomath
\begin{align}
x[i]=
&\mathrm{floor}
\left[
x_\mathrm{max}\frac{A_0\cos(\omega t[i]+\theta_0)-A_0\omega j(t[i])\sin(\omega t[i]+\theta_0)}{A_\mathrm{R}}\right.\nonumber\\
&\left.\quad\frac{+A_\mathrm{M}(t[i])\cos(\omega t[i]+\theta_0)+n_\mathrm{PI}(t[i])+a_\mathrm{total}(t[i])}{A_\mathrm{R}}\right], \label{e30-1}
\end{align}
where the amplitudes and bandwidths of $j(t[i])$, $A_\mathrm{M}(t[i])$, $n_\mathrm{PI}(t[i])$, and $a_\mathrm{total}(t[i])$ are selected arbitrarily. The frequency of the pure sinusoidal wave $f_\mathrm{C}=\omega/2\pi $ was varied between $f_\mathrm{P}/4-120~\mathrm{Hz}\leqslant f_\mathrm{C}\leqslant f_\mathrm{P}/4+120$ Hz, where $f_\mathrm{P}= 48~\mathrm{kHz}$ is the sampling frequency of the playback device, to verify the accuracy of the program. The time interval between $t[i+1]$ and $t[i]$ was set to $f_\mathrm{R}^{-1}\approx 5.2083~\rmu\mathrm{s}$, where $f_\mathrm{R}=192~\mathrm{kHz}$ is the sampling rate of the recorders. Herein, we show the results obtained under the three conditions outlined below: (i) $\mathrm{dev}\{j(t[i])\}\ne 0$ and $A_\mathrm{M}(t)=n_\mathrm{PI}(t)=a_\mathrm{total}(t)=0$, (ii) $\mathrm{dev}\{A_\mathrm{M}(t[i])\}\ne 0$ and $j(t)=n_\mathrm{PI}(t)=a_\mathrm{total}(t)=0$, and (iii) $\mathrm{dev}\{n_\mathrm{PI}(t[i])\}\ne 0$ and $j(t)=A_\mathrm{M}(t)=a_\mathrm{total}(t)=0$. Using these dummy waveforms, we confirm that outcomes of the program are accurate.

\subsection{Creation of dummy waveforms}
First, we create artificial jitter, AM and PI noise without bandwidth limitation, which are represented as $j_\mathrm{FBW}(t[i])$, $A_\mathrm{M,FBW}(t[i])$, and $n_\mathrm{PI,FBW}(t[i])$, respectively. These are expressed as follows: 
\linenomath
\begin{align}
j_\mathrm{FBW}(t[i])&=J\cdot\mathrm{randn}[i],\\
A_\mathrm{M,FBW}(t[i])&=A_0\omega J\cdot\mathrm{randn}[i],\\
n_\mathrm{PI,FBW}(t[i])&=A_0\omega J\cdot\mathrm{randn}[i],
\end{align}
where $\mathrm{randn}[~]$ is a MATLAB function that generates normally distributed random numbers with a standard deviation of 1.The carrier frequency is set to $\omega/2\pi =11.884~877~\mathrm{kHz}$. The carrier amplitude is set to $A_0=0.9 A_\mathrm{R}=0.9~\mathrm{FS}$, where $\mathrm{FS}$ denotes the full scale of the recorder input. The amplitude of jitter, represented by $J$, is set to $J=160~\mathrm{ps}$. Therefore, the deviation of the artificial jitter becomes $\mathrm{dev}\{j_\mathrm{FBW}(t[i])\}=160~\mathrm{ps}$. The bandwidth of $j_\mathrm{FBW}(t[i])$ is $f_\mathrm{R}/2=96~\mathrm{kHz}$ because of the sampling theorem.

Second, we apply bandwidth limitation to $j_\mathrm{FBW}(t[i])$, $A_\mathrm{M,FBW}(t[i])$, and $n_\mathrm{PI,FBW}(t[i])$, and obtain artificial jitter $j(t[i])$, AM $A_\mathrm{M}(t[i])$ and PI noise $n_\mathrm{PI}(t[i])$. The applied bandwidth limitations are low pass filter (LPF) and band pass filter (BPF) types, represented as 
\linenomath
\begin{align}
j(t[i])&=\mathcal{LF}\{j_\mathrm{FBW}(t[i])\} \label{Eq:j_FBW}\\
A_\mathrm{M}(t[i])&=\mathcal{LF}\{A_\mathrm{M,FBW}(t[i])\} \label{Eq:AM_FBW}\\
n_\mathrm{PI}(t[i])&=\mathcal{BPF}\{n_\mathrm{PI,FBW}(t[i])\} \label{Eq:PI_FBW}
\end{align}
where $\mathcal{LF}\{~\}$ denotes the frequency component of $0\leqslant f\leqslant B_\mathrm{W}$, and $\mathcal{BPF}\{~\}$ denotes the frequency component of $f_\mathrm{C}-B_\mathrm{W}\leqslant f\leqslant f_\mathrm{C}+B_\mathrm{W}$. We set the cut-off frequency of the LPF and the half width of the BPF as $B_\mathrm{W}=6~\mathrm{kHz}$, which is approximately $f_\mathrm{C}/2$. Because of this bandwidth limitation, the deviation of the artificial jitter is reduced to $\mathrm{dev}\{j(t[i])\}=J\{B_\mathrm{W}/(f_\mathrm{R}/2)\}^{1/2}=40~\mathrm{ps}$. Similarly, the deviation of AM and PI noise becomes $(\omega A_0)^{-1}\mathrm{dev}\{A_\mathrm{M}(t[i])\} \approx 40~\mathrm{ps}$, and $(\omega A_0)^{-1}\mathrm{dev}\{n_\mathrm{PI}(t[i])\} \approx 40\sqrt{2}~\mathrm{ps}$.

Finally, we set the parameters in Eq. (\ref{e30-1}) as $x_\mathrm{max}=2^{23}-1$, and $\theta_0=0$, respectively. Consequently, dummy waveforms comprising (i) a pure sinusoidal wave and jitter, (ii) a pure sinusoidal wave and AM, and (iii) a pure sinusoidal wave and PI noise are created. The dummy waveforms were saved as a 24-bit wave files and then loaded.

\subsection{FDA of dummy waveforms}
\begin{figure}
\figcolumn{\fig{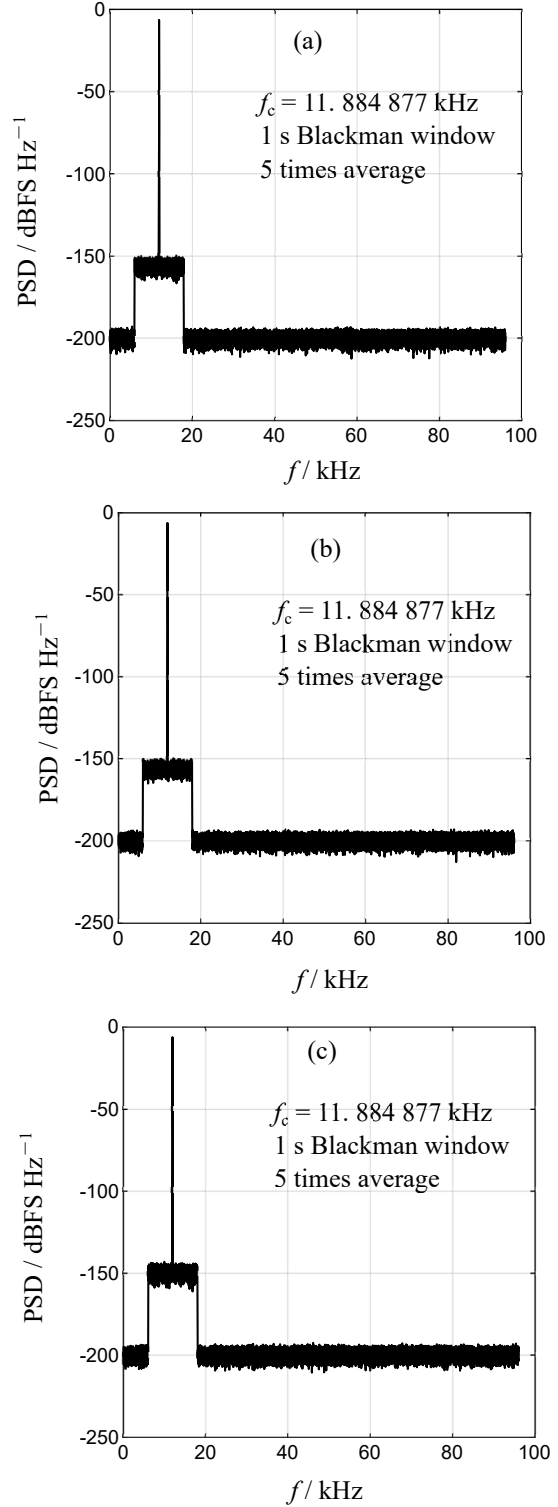}{0.4\figwidth}{}}
\caption{\label{Fig:spectrum}{The outcomes of the FDA. (a) PSD of (i). (b) PSD of (ii). (c) PSD of (iii).}}
\end{figure}

We demonstrate the spectrum of the dummy waveforms (i), (ii) and (iii) in \autoref{Fig:spectrum}(a), (b) and (c), respectively. 
The horizontal axis represents frequency $f$, and the vertical axis represents power spectral density (PSD) in $\mathrm{dBFS}~\mathrm{Hz}^{-1}$, where FS denotes full scale.

The result of FDA with (i) is shown in \autoref{Fig:spectrum}(a). A peak is seen at frequency $f\approx 12~\mathrm{kHz}$. The height and width are $-5~\mathrm{dBFS~Hz^{-1}}$ and $1~\mathrm{Hz}$, respectively. This peak corresponds to PSD of the pure sinusoidal wave; the area of the peak is approximately $(10^{-5/10}~\mathrm{FS^2~Hz^{-1}})(1~\mathrm{Hz})= 10^{-5/10}~\mathrm{FS^2}$. The relative power of carrier wave absorbed by the recorder is expressed as follows 
\linenomath
\begin{align}
P_\mathrm{C}/\mathrm{FS}^2 &=\frac{\langle\{F_\mathrm{pure}(t)\}^2\rangle}{A_\mathrm{R}^2}=\frac{(A_0/A_\mathrm{R})^2}{2},
\end{align}
where the symbol $\langle~\rangle$ represents temporal average. Hence, $P_\mathrm{C}$ is calculated as $P_\mathrm{C}=(0.9~\mathrm{FS})^2/2~\approx 10^{-4/10}~\mathrm{FS^2}$. This value corresponds to the area of the peak. White noise floor is seen around the level of $-200~\mathrm{dBFS}~\mathrm{Hz}^{-1}$, which corresponds to quantization noise; the area of the noise floor is calculated as $(10^{-200/10}~\mathrm{FS}^2~\mathrm{Hz}^{-1})(f_\mathrm{R}/2)\approx 10^{-150/10}~\mathrm{FS^2}$. The theoretical S/N of quantization noise for $N_\mathrm{R}=24~ \mathrm{bit}$ ADC is computed as $6.02 N_\mathrm{R}+1.76~\mathrm{dB}=146~\mathrm{dBc}$\citep{Kester09}. Hence, the relative power of the quantization noise is $10^{-146/10}\{(1~\mathrm{FS})^2/2\}=10^{-149/10}~\mathrm{FS^2}$. This value is close to the area of the white noise floor seen in \autoref{Fig:spectrum}(a). A rectangular spectrum is seen at $6~\mathrm{kHz} \leqslant f \leqslant 18~\mathrm{kHz}$. The top level is $-155~\mathrm{dBFS~Hz^{-1}}$. This rectangular spectrum corresponds to PSD of the artificial jitter; the area of the rectangular spectrum is calculated as $(10^{-155/10}~\mathrm{FS^2~Hz^{-1}}-10^{-195/10}~\mathrm{FS^2~Hz^{-1}})(18~\mathrm{kHz}-6~\mathrm{kHz})=10^{-114/10}~\mathrm{FS^2}$. Based on Eq. (\ref{e30-1}), the relative power of jitter is written as
\linenomath
\begin{align}
P_\mathrm{jitter}/\mathrm{FS}^2 &=\frac{\langle\{n_\mathrm{jitter}(t)\}^2\rangle}{A_\mathrm{R}^2}= \frac{(\omega A_0/A_R)^2\mathbb{V} \{j(t[i])\}}{2}. \label{Eq:P_jitter}
\end{align}
Hence, $P_\mathrm{jitter}$ is calculated as $P_\mathrm{jitter}\approx 10^{-114/10}~\mathrm{FS^2}$. This value is close to the area of the rectangular spectrum seen in \autoref{Fig:spectrum}(a). 

The results of FDA with (ii) are demonstrated in \autoref{Fig:spectrum}(b). This resembles \autoref{Fig:spectrum}(a) and can be interpreted similarly. A peak is observed at $f\approx 12~\mathrm{kHz}$, which corresponds to the PSD of the pure sinusoidal wave. The white noise floor is observed at the level of $-195~\mathrm{dBFS~Hz^{-1}}$, which corresponds to quantization noise. A rectangular spectrum is observed at at $6~\mathrm{kHz} \leqslant f \leqslant 18~\mathrm{kHz}$. The top level is $-155~\mathrm{dBFS~Hz^{-1}}$. This rectangular spectrum corresponds to the PSD of artificial AM; the area of the rectangular spectrum is calculated as $10^{-114/10}~\mathrm{FS^2}$. Based on Eq. (\ref{e30-1}), the relative power of AM is written as 
\linenomath
\begin{align}
P_\mathrm{AM}/\mathrm{FS}^2 &=\frac{\langle\{n_\mathrm{AM}(t)\}^2\rangle}{A_\mathrm{R}^2}= \frac{\mathbb{V}\{A_\mathrm{M}(t[i])\}}{2A_\mathrm{R}^2}. \label{Eq:P_AM}
\end{align}
Hence, $P_\mathrm{AM}$ is computed as $P_\mathrm{AM}\approx 10^{-114/10}~\mathrm{FS^2}$. This value is close to the area of the rectangular spectrum seen in \autoref{Fig:spectrum}(b).

As can be observed, the PSD of (i) and (ii) are indistinguishable. This was caused by the conditions of Eq. (\ref{Eq:j_FBW}) and (\ref{Eq:AM_FBW}), and was expected. We chose these conditions because these outcomes clearly show the disadvantages of the FDA; jitter and AM are indistinguishable from FDA. As we described in the introduction section, TDA, e.g., Hirbert transform analysis\citep{Nishimura10} (HTA) and ZCA, enables us to eliminate AM. 

The outcomes of FDA with (iii) are demonstrated in \autoref{Fig:spectrum}(c). The difference from \autoref{Fig:spectrum}(a) is that the top level of the rectangular spectrum is changed to $-150~\mathrm{dBFS~Hz^{-1}}$. This angular spectrum corresponds to PSD of artificial PI noise; the area of the rectangular spectrum is computed as $10^{-109/10}~\mathrm{FS^2}$. Based on Eq. (\ref{e30-1}), the relative power of PI noise is expressed as
\linenomath
\begin{align}
P_\mathrm{PI}/\mathrm{FS}^2 &=\frac{\langle\{n_\mathrm{PI}(t)\}^2\rangle}{A_\mathrm{R}^2}= \frac{\mathbb{V}\{n_\mathrm{PI}(t[i])\}}{A_\mathrm{R}^2} \label{Eq:P_PI}.
\end{align}
Hence, $P_\mathrm{PI}$ is calculated as  $P_\mathrm{PI}\approx10^{-108/10}~\mathrm{FS^2}$. This value is close to the area of the rectangular spectrum seen in \autoref{Fig:spectrum}(c). This indicates that FDA enables the measurement of the total noise $(P_\mathrm{jitter}+P_\mathrm{AM}+P_\mathrm{PI})/P_\mathrm{C}$. Nevertheless, because $P_\mathrm{AM}$ is not eliminated, the phase noise. i.e., $(P_\mathrm{jitter}+P_\mathrm{PI})/P_\mathrm{C}$, is overestimated with FDA. Therefore, TDA is required for the absolute measurement of sampling jitter.

\subsection{HTA of dummy waveform}
\begin{figure}
\figcolumn{\fig{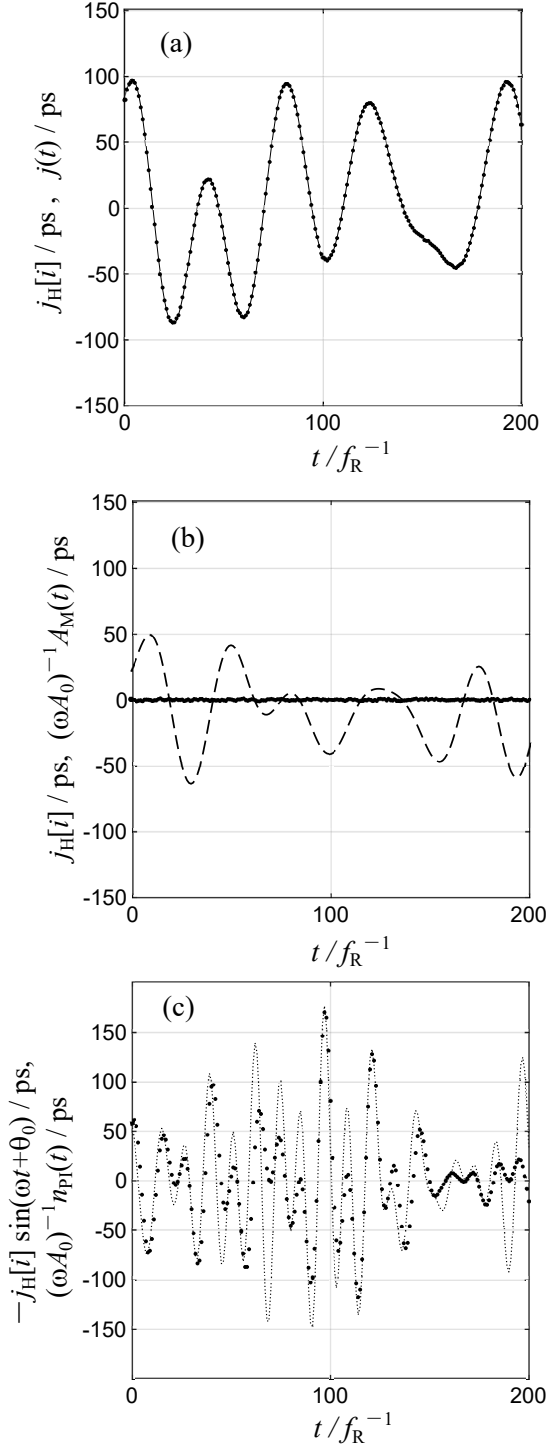}{0.4\figwidth}{}}
\caption{\label{Figure7}{The results of the HTA. (a) Dots represent $j_\mathrm{H}[i]$ obtained from (i). The solid line represents $j(t)$. (b) Dots represents $j_\mathrm{H}[i]$ obtained from (ii). The dashed line represents $A_\mathrm{M}(t)$. (c) Dots represent $j_\mathrm{H}[i]$ obtained from (iii). The dotted line represents $n_\mathrm{PI}(t)$.}}
\end{figure}

In this subsection, we consider the HTA\citep{Nishimura10} as an existing technique of TDA. The dummy waveform (i) is analyzed using HTA. The outcome is illustrated in \autoref{Figure7}(a). 
The horizontal axis represents time $t$. The points $j_\mathrm{H}[i]$ obtained from the dummy waveform (i) are plotted at $t=(i-1)f_\mathrm{R}^{-1}$ using dot markers. Artificial jitter $j(t)$ is drawn as a solid line for comparison. As shown, HTA successfully enables us to sample $j(t)$ with a sampling rate of $f_\mathrm{R}$. 

The result of the HTA with the dummy waveform (ii) is demonstrated in \autoref{Figure7}(b). The obtained jitter $j_\mathrm{H}[i]$ is plotted at $t=(i-1)f_\mathrm{R}^{-1}$ using dot markers. For comparison, artificial AM $A_\mathrm{M}(t)$ is represented as a dashed line. As can be seen, the data points of the HTA obtained from (ii) are nearly zero and not affected by AM. This is the advantage of the TDA compared with the FDA; jitter can be observed, whereas AM is eliminated.

HTA results with the dummy waveform (iii) is shown in \autoref{Figure7}(c). The up-converted jitter $-j_\mathrm{H}[i]\sin(\omega t+\theta_0)$ is plotted at $t=(i-1)f_\mathrm{R}^{-1}$ using dot markers. For comparison, artificial PI noise $n_\mathrm{PI}(t)$ is drawn as a dotted line. The up conversion of $j_\mathrm{H}[i]$ is necessary based on Eq. (4). As can be observed, HTA fails to sample $n_\mathrm{PI}(t)$. The phase noise $(P_\mathrm{jitter}+P_\mathrm{PI})/P_\mathrm{C}$ is underestimated with HTA. Therefore, an improved TDA better than HTA is essential for the absolute measurement of sampling jitter.

\subsection{ZCA of dummy waveforms}
\begin{figure}
\figcolumn{\fig{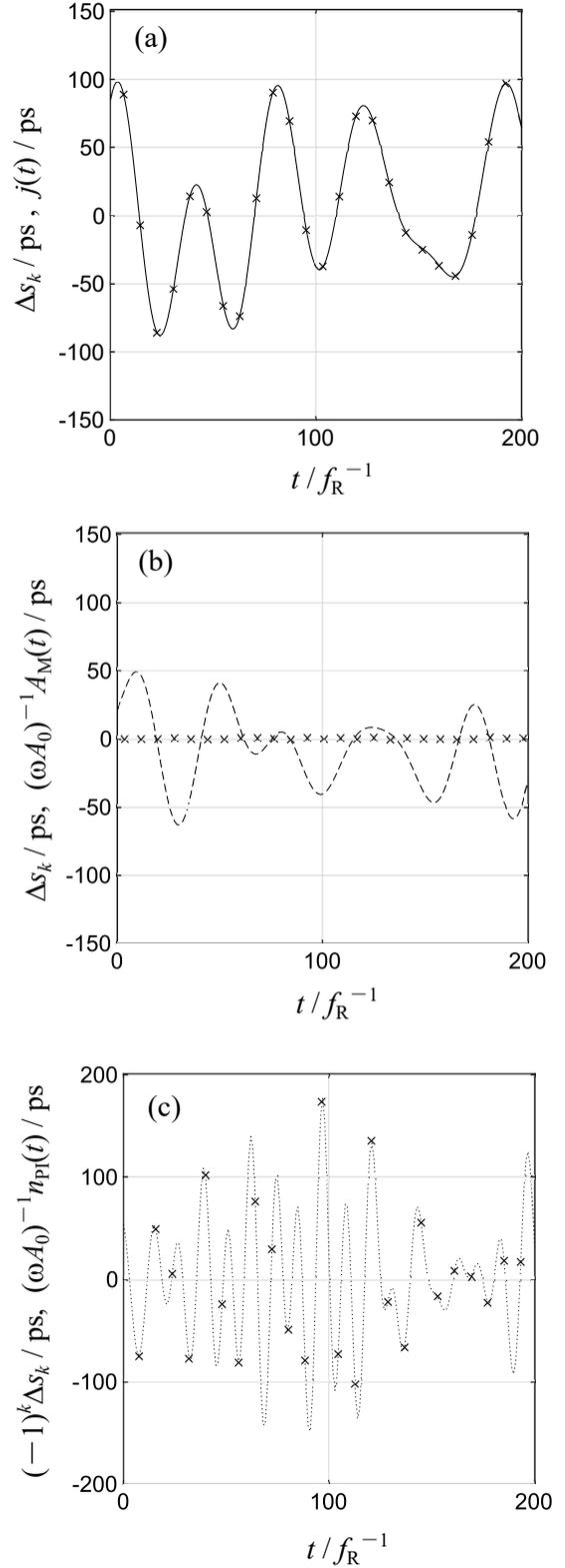}{0.4\figwidth}{}}
\caption{\label{Fig:ZCA1}{(a) ZCFs $\varDelta s_k$ (crosses) gotten from (i). The solid line depicts $j(t)$. (b) ZCFs $\varDelta s_k$ (crosses) obtained from (ii). The dashed line represents $A_\mathrm{M}(t)$. (c) ZCFs $\varDelta s_k$ (crosses) acquired from (iii). The dotted line represents $n_\mathrm{PI}(t)$.}}
\end{figure}

In this subsection, we consider the ZCA as an improved TDA. The dummy waveform (i) is analyzed using the developed ZCA program. The result is shown in \autoref{Fig:ZCA1}(a). 
The horizontal axis represents time $t$. The obtained ZCFs $\varDelta s_k$ are plotted at $t=s'_k$ using cross markers. For comparison, $j(t)$ is drawn as a solid line. As shown in \autoref{Fig:ZCA1}(a), ZCA successfully enables us to sample $j(t)$ at the ideal timing of $t=s'_k$. The quite difference from \autoref{Figure7}(a) is that the data points appear at $t=s'_k$, i.e., the zero-crossing timing of the carrier wave; the sampling rate of ZCA is $2f_\mathrm{C}$, as noted in Section {II D}. This is sufficiently high to reconstruct $j(t)$ from $\varDelta s_k$, as guaranteed by the sampling theorem. 

The result of ZCA with the dummy waveform (ii) is shown in \autoref{Fig:ZCA1}(b). The obtained ZCFs $\varDelta s_k$ are plotted at $t=s'_k$ using cross markers. For comparison, $A_\mathrm{M}(t)$ is drawn as a dashed line. As can be seen, ZCFs are nearly zero and not affected by AM. This is natural from Eq. (21); jitter and PI noise can be observed, whereas AM is eliminated. From the ZCA results of (i) and (ii), we can say that the accuracy of the ZCA program which we developed is adequate in the region of $10~\mathrm{ps}$ order.

The result of the ZCA with the dummy waveform (iii) is demonstrated in \autoref{Fig:ZCA1}(c). The obtained ZCFs $(-1)^k\varDelta s_k$ are plotted at $t=s'_k$ using cross markers. The factor $(-1)^k$ is necessary based on Eq. (21). For comparison, $n_\mathrm{PI}(t)$ is drawn as a dotted line. As can be seen, ZCA successfully enables us to sample $n_\mathrm{PI}(t)$. This indicates that the phase noise $(P_\mathrm{jitter}+P_\mathrm{PI})/P_\mathrm{C}$ is suitably estimated using ZCA.

\bibliography{bib00.bib}

\begin{thebibliography}{10}
\def\enquote#1,{``#1,''}
\def\enxquote#1{``#1''}
\expandafter\ifx\csname url\endcsname\relax
  \def\url#1{\texttt{#1}}\fi
\expandafter\ifx\csname urlprefix\endcsname\relax\def\urlprefix{URL }\fi
\providecommand{\bibinfo}[2]{#2}
\def\plainquote#1{``#1''}
\providecommand{\noopsort}[1]{}
\providecommand{\switchargs}[2]{#2#1}
\providecommand{\dourl}[1]{\href{http://#1}{\nolinkurl{#1}}}
  \def\eatspace #1{#1}

\bibitem{Dunn92}
\bibinfo{author}{J.~Dunn}, \enquote{\bibinfo{title}{Jitter: Specification and
  assessment in digital audio equipment}}, in
  \emph{\bibinfo{booktitle}{Proceedings of the 93rd AES Convention}},
  \bibinfo{address}{San Francisco, CA} (\bibinfo{year}{October 1--4, 1992}), p.
  \bibinfo{pages}{3361}.

\bibitem{Dunn94Feb}
\bibinfo{author}{J.~Dunn} and \bibinfo{author}{I.~Dennis},
  \enquote{\bibinfo{title}{The diagnosis and solution of jitter-related
  problems in digital audio systems}}, in \emph{\bibinfo{booktitle}{Proceedings
  of the 96th AES Convention}}, \bibinfo{address}{Amsterdam, The Netherlands}
  (\bibinfo{year}{February 26--March 1, 1994}), p. \bibinfo{pages}{3868}.

\bibitem{Dunn94May}
\bibinfo{author}{J.~Dunn}, \enquote{\bibinfo{title}{The diagnosis and solution
  of jitter-related problems in digital audio systems}}, in
  \emph{\bibinfo{booktitle}{Proceedings of the AES 9th UK Conference: Managing
  the Bit Budget (MBB)}}, \bibinfo{address}{London, UK} (\bibinfo{year}{May
  16--17, 1994}), pp. \bibinfo{pages}{148--166}.

\bibitem{Dunn00}
\bibinfo{author}{J.~Dunn}, \enquote{\bibinfo{title}{Jitter theory}},
  \bibinfo{type}{Audio Precision TECHNOTE TN-23}, \bibinfo{institution}{Audio
  Precision}, \bibinfo{address}{Beaverton, Oregon} (\bibinfo{year}{2000}).

\bibitem{Nishimura10}
\bibinfo{author}{A.~Nishimura} and \bibinfo{author}{N.~Koizumi},
  \enquote{\bibinfo{title}{Measurement of sampling jitter in analog-to-digital
  and digital-to-analog converters using analytic signals}},
  \bibinfo{journal}{Acoustical Science and Technology} \textbf{31}(2),
  \bibinfo{pages}{172--180} (\bibinfo{year}{2010}).

\bibitem{MacLean40}
\bibinfo{author}{W.~R. MacLean}, \enquote{\bibinfo{title}{Absolute measurement
  of sound without a primary standard}},  \bibinfo{journal}{The Journal of the
  Acoustical Society of America} \textbf{12}, \bibinfo{pages}{140--146}
  (\bibinfo{year}{1940}).

\bibitem{Barrera-Figueroa18}
\bibinfo{author}{S.~Barrera-Figueroa}, \enquote{\bibinfo{title}{Free-field
  reciprocity calibration of measurement microphones at frequencies up to 150
  khz}},  \bibinfo{journal}{The Journal of the Acoustical Society of America}
  \textbf{144}(4), \bibinfo{pages}{2575--2583} (\bibinfo{year}{2018}).

\bibitem{Rubiola10}
\bibinfo{author}{E.~Rubiola} and \bibinfo{author}{F.~Vernotte},
  \enquote{\bibinfo{title}{The cross-spectrum experimental method}},
  (\bibinfo{year}{2010}) \dourl{arXiv:physics.ins-det/1003.0113}.

\bibitem{Ashihara05}
\bibinfo{author}{K.~Ashihara}, \bibinfo{author}{S.~Kiryu},
  \bibinfo{author}{N.~Koizumi}, \bibinfo{author}{A.~Nishimura},
  \bibinfo{author}{J.~Ohga}, \bibinfo{author}{M.~Sawaguchi}, and
  \bibinfo{author}{S.~Yoshikawa}, \enquote{\bibinfo{title}{Detection threshold
  for distortions due to jitter on digital audio}},
  \bibinfo{journal}{Acoustical Science and Technology} \textbf{26}(1),
  \bibinfo{pages}{50--54} (\bibinfo{year}{2005}).

\bibitem{Melchior19}
\bibinfo{author}{V.~R. Melchior}, \enquote{\bibinfo{title}{High resolution
  audio: A history and perspective}},  \bibinfo{journal}{Journal of the Audio
  Engineering Society} \textbf{67}(5), \bibinfo{pages}{246--257}
  (\bibinfo{year}{2019}).

\bibitem{Nittono20}
\bibinfo{author}{H.~Nittono}, \enquote{\bibinfo{title}{High-frequency sound
  components of high-resolution audio are not detected in auditory sensory
  memory}},  \bibinfo{journal}{Scientific Reports} \textbf{10},
  \bibinfo{pages}{21740} (\bibinfo{year}{2020}).

\bibitem{Note1}
\bibinfo{note}{The player yields a sinusoidal wave of $f_\protect \mathrm
  {C}=f_\protect \mathrm {P}/4$; however, the frequency measured by the
  recorder is not exactly $f_\protect \mathrm {P}/4$.}

\bibitem{Note2}
\bibinfo{note}{See Appendix for the results of numerical simulation}.

\bibitem{FFTWsite}
\bibinfo{author}{M.~Frigo} and \bibinfo{author}{S.~G. Johnson},
  \enquote{\bibinfo{title}{{FFTW} (version 3.3.10) [computer program]}},
  (\bibinfo{year}{2005}) \bibinfo{note}{\url{http://www.fftw.org} (Last viewed
  April 27, 2022)}.

\bibitem{Feldhaus16Apr}
\bibinfo{author}{G.~Feldhaus} and \bibinfo{author}{A.~Roth},
  \enquote{\bibinfo{title}{A 1 {MH}z to 50 {GH}z direct down-conversion phase
  noise analyzer with cross-correlation}},  \bibinfo{journal}{2016 European
  Frequency and Time Forum (EFTF)} \bibinfo{pages}{1--4} (\bibinfo{year}{2016})
  \dodoi{10.1109/EFTF.2016.7477759}.

\bibitem{Feldhaus16AN}
\bibinfo{author}{G.~Feldhaus}, \bibinfo{author}{G.~Roesel},
  \bibinfo{author}{A.~Roth}, and \bibinfo{author}{J.~Wolle},
  \enquote{\bibinfo{title}{Measurement uncertainty analysis and traceability
  for phase noise}}, \bibinfo{type}{Application Note No. 1EF95},
  \bibinfo{institution}{Rohde \& Shwarz}, \bibinfo{address}{Munich, Germany}
  (\bibinfo{year}{2016}).

\bibitem{Kester09}
\bibinfo{author}{W.~Kester}, \enquote{\bibinfo{title}{Taking the mystery out of
  the infamous formula, ``{SNR} = 6.02{{\it N}} + 1.76 d{B},'' and why you
  should care}}, \bibinfo{type}{Analog Devices MT-001 TUTORIAL},
  \bibinfo{institution}{Analog Devices} (\bibinfo{year}{2009}).

\end{thebibliography}
\end{document}